\documentclass{aa}
\usepackage[varg]{txfonts}
\usepackage[T1]{fontenc}
\usepackage[utf8]{inputenc}
\usepackage{textcomp} 
\usepackage{gensymb}
\usepackage{natbib}
\usepackage{amsmath}
\usepackage{graphicx,subfig}
\usepackage{color}

\usepackage[normalem]{ulem}

\fancypagestyle{firstpage}{%
	\fancyhf{}%
	\fancyfoot[RO]{\thepage}%
	\fancyfoot[LE]{\thepage}%
}

\begin{document}

\titlerunning{Resonance captures in binaries}

\title{Resonance capture and long-term evolution of planets in binary star systems}
\author{A. Roisin\inst{1} \and N. Doukhanin\inst{1} \and J. Teyssandier\inst{1}
	\and A.-S. Libert\inst{1}
}
\institute{naXys, Department of Mathematics, University of Namur, 61 Rue de Bruxelles, B-5000 Namur, Belgium}
\date{Received ... / Accepted ...}
\abstract {} {The growing population of planets discovered in orbit around one stellar component of a binary star raises the question of the influence of the binary companion on the formation process of planetary systems. The aim of this work is to study the impact of a binary companion on the evolution of two-planet systems during both the type-II migration phase and their long-term evolution after the dissipation of the protoplanetary disk.} 
{We use the symplectic integrator SyMBA, modified to include a wide binary companion. We also include the Type-II migration of giant planets during the protoplanetary disk phase with suitable eccentricity and inclination damping as well as the gravitational potential acting on the planets due to the disk and the nodal precession of the disk induced by the binary companion. We consider various inclinations, eccentricities, and separations of the binary companion.}
{Disk migration allows the formation of planet pairs in mean-motion resonances despite the presence of the binary companion. When the binary separation is wide (1000~au), the timescale of the perturbations it raises on the planets is longer than the disk's lifetime and resonant pairs are routinely formed in the 2:1, 5:2 and 3:1 commensurabilities. Provided the planet-planet interaction timescale is smaller than  the binary perturbations timescale, these systems can remain in resonance long after the disk has dissipated. When the binary separation is smaller (250 au), only planets in the 2:1 resonance tend to remain in a resonant state and more chaotic evolutions are observed, as well as more ejections. After those ejections, the remaining planet can become eccentric due to the perturbations from the binary companion and for strongly inclined binary companions captures in the von Ziepel-Lidov-Kozai resonance can occur, while in systems with two planets this mechanism is quenched by planet-planet interactions. Our simulations reveal that the interplay between planet-disk, planet-planet and planet-binary interactions can lead to the formation of resonant pairs of planets which remain stable over timescales much longer than the disk's lifetime.}
{}


\maketitle

\section{Introduction}
Planetary system formation under the influence of a binary companion is an important question since it is estimated that about half of the Sun-like stars are part of multiple star systems \citep{Duquennoy_1991,Raghavan_2010}. Until now, more than 100 S-type planets (also called circumprimary planets) have been discovered and their eccentricities are slightly higher than around single star systems \citep{Kaib_2013}.

A binary companion plays an important role in the formation process. Firstly, it has a considerable impact on the protoplanetary disk. Regarding the disk formation, it was observed that disks in close binary star systems are less present and if they are, they tend to be less massive than in single star systems \citep{Kraus_2012,Franchini_2019,Chachan_2019}. The binary companion can induce a truncation of the disk, as shown by \cite{Artymowicz_1994,Savonije_1994}.  The binary companion also influences the shape of the disk \citep{Terquem_1993,Papaloizou_1995} and could even lead to the warping of the disk if the self gravity or pressure forces are not strong enough to maintain it uniform \citep{Batygin_2011,Zanazzi_2017}. Furthermore, the disk nodally precesses due to the presence of the binary companion \citep{Batygin_2011, Roisin_2021b}.

Secondly, the binary companion also influences the planet embedded in the protoplanetary disk. The planet-disk interactions were extensively studied for close binaries by means of 3D hydrodynamical simulations where a decoupling between the motion of the planet and the disk was generally observed \citep[e.g.,][]{Xiang_2014, Picogna_2015,Lubow_2016,Martin_2016}. Regarding wide binaries, it was shown that the disk gravitational potential acting on the planet and the damping forces exerted by the disk on the planet tend to keep the latter in the midplane of the former \citep[e.g.,][]{Roisin_2021b}. 

One key effect affecting the planets in binary star systems is the von Zeipel-Lidov-Kozai (ZLK) resonance which consists of coupled eccentricity and inclination variations for the planets \citep{vonZeipel_1910, Lidov_1962, Kozai_1962}. This effect influences the formation of giant planets in binary stars in multiple ways. It has been shown that the binary companion can generate ZLK cycles in the disk \citep{Martin_2014, Fu_2015a} which could be suppressed by the self-gravity of the disk \citep{Batygin_2011,Rafikov_2015_b}. Moreover, for sufficiently inclined disks, ZLK instabilities can arise \citep{Lubow_2017,Zanazzi_2017}. Inclined binary companions also influence planets migrating in the protoplanetary disk via the ZLK effect \citep{Roisin_2021a}, but this effect is counteracted by the gravitational potential of the disk acting on the planets in the case of wide binaries \citep{Roisin_2021b}. Finally, it has been widely shown that the ZLK mechanism, when associated with tidal friction, could lead to the formation of hot Jupiters in binary star systems \citep[e.g.,][]{Fabrycky_2007}.

Most of the planet formation studies focus on close binary systems, due to the strong interaction between the planets and the binary companion. However a wide binary could also in some case have a significant impact on the S-type planetary system architecture. This work aims at extending the studies on single planet systems made by \cite{Roisin_2021a} and \cite{Roisin_2021b} by considering the influence of a wide binary companion on the resonance captures of two giant planets migrating in the Type-II regime during the protoplanetary disk phase where the gravitational influence of the disk on the planets and the mutual gravitational influence between the planets also act. The long-term evolution of the systems after the dispersal of the disk will be thoroughly investigated. We will also focus our attention on the ZLK resonance with the wide binary companion and how it could play a role during the two stages mentioned above.

The paper is organized as follows. Section~\ref{sec:methods} describes the set-up of the $N$-body simulations carried out in this work. Section~\ref{sec:timescales} presents an overview of the relevant timescales of the problem. Section~\ref{sec:wide_bin} presents the results of the simulations for a wide binary companion at 1000~au by outlining the system architectures as well as the different dynamical effects observed during and after the disk phase,  while a similar analysis is performed in Section~\ref{sec:close_bin} for closer binary stars. The long-term stability of the systems found in our simulations is assessed in Section~\ref{sec:stability}. In Section~\ref{sec:obs}, we discuss our results in light of the observational data on S-type planets. Finally our conclusions are given in Section~\ref{sec:conclusion}.

\section{Methods}
\label{sec:methods}
In this work, we follow the evolution of two planets perturbed by a binary companion during and after the disk phase. Different semi-major axis $a_B$, eccentricity $e_B$, and inclination $i_B$ values are considered for the binary. The masses of both stars are $1~M_{\odot}$ in all the simulations.

To run those simulations, the code used in this work is the well-known $N$-body SyMBA integrator, which handles close encounters \citep{Duncan_1998}. It was modified to include a wide binary companion following \cite{Chambers_2002} \citep[see][for more details]{Roisin_2021a}. The code also considers the gravitational influence of the protoplanetary disk in which the planets are embedded, namely the type-II migration of the planets \citep{Ivanov_1999,Nelson_2000,Crida_2007}, through the acceleration  \citep{Papaloizou_2000}
\begin{equation}
\mathbf{a}_{\rm mig} = - \frac{\mathbf{v}_{\rm pl}}{\tau_{\rm mig}},
\end{equation}
with $\mathbf{v}_{\rm pl}$ the velocity of the migrating planet and $\tau_{\rm mig}$ the timescale for the Type-II regime given by \citep[both in the disk-dominated case and planet-dominated cases, see][for more details]{Sotiriadis_2017}: 
\begin{equation}
\tau_{\rm mig} = \frac{2}{3} \alpha^{-1} h^{-2} \Omega_{\rm pl}^{-1}  \times \max\left\{1, \frac{m_{\rm pl}}{ (4 \pi/3) \Sigma(r_{\rm pl}) r_{\rm pl}^2}\right\},
\end{equation}
with $\alpha=0.005$ the classical value for the Shakura-Sunyanev viscosity parameter \citep{Shakura_1973}, $h=0.05$  the disk aspect ratio, $\Omega_{\rm pl}^{-1}$ the orbital frequency of the planet, $a_{\rm pl}$ the semi-major axis of the planet, $m_{\rm pl}$ the mass of the planet, $r_{\rm pl}$ the distance of the planet to the host star, and $\Sigma \propto r^{-0.5}$ the surface density profile of the disk. Unless otherwise stated, in the simulations, we fix the disk inner and outer edges to $a_{\rm in}=0.05$ AU and $a_{\rm out}=30$ AU. The code also adopts a smooth transition in the gas-free inner cavity using an hyperbolic tangent function $\tanh \left(\frac{r-R_{\rm in}}{\Delta r}\right)$ where $\Delta r=0.001$ AU, following \cite{Matsumoto_2012}.

Moreover, the inclination and eccentricity damping induced on the planets by the disk is included through the formulas deduced from the 3D hydrodynamical simulations in \cite{Bitsch_2013}: 
\begin{multline}
\frac{de}{dt}(m_{\rm pl}, e_{\rm pl}, i_{\rm pl}) = - \frac{m_{\rm ld}}{0.025 m_A} \left(a_{e} \left[i_{\rm pl} + \frac{m_{\rm pl}}{3}\right]^{-2b_{e}} + c_{e} i_{\rm pl}^{-2d_{e}}\right)^{-\frac{1}{2}} \\ \hspace{2cm} + 12.65 \frac{m_{\rm pl} m_{\rm ld}}{m_A^2} e _{\rm pl} \exp\left(-\left[\frac{i_{\rm pl}}{m_{\rm pl}}\right]^2 \right),\\
\frac{di}{dt}(m_{\rm pl}, e_{\rm pl}, i_{\rm pl}) = - \frac{m_{\rm ld}}{0.025 m_A} \left(a_{i} i_{\rm pl}^{-2 b_{i}} \exp \left(\frac{-(i_{\rm pl}/g_{i})^2}{2}\right) \right.\\ \left.+ c_{i} \left[\frac{i_{\rm pl}}{40}\right]^{-2d_{i}}\right)^{-\frac{1}{2}},
\label{e_dot}
\end{multline}
with $e_{\rm pl}$ and $i_{\rm pl}$ the planetary eccentricity and inclination, respectively, $m_A$ the mass of the central star, and with the coefficients defined as
\begin{equation}
a_{e} = 80 e_{\rm pl}^{-2} \exp\left\{-e_{\rm pl}^2 \frac{m_{\rm pl}}{0.26}\right\} 15^{m_{\rm pl}}  \left(20+11m_{\rm pl}-m_{\rm pl}^2\right),
\end{equation}
\begin{equation}
b_{e} = 0.3 m_{\rm pl},
\end{equation}
\begin{equation}
c_{e} = 450 + 2 ^{m_{\rm pl}},
\end{equation}
\begin{equation}
d_{e} = -1.4 + \frac{\sqrt{m_{\rm pl}}}{6}.
\end{equation}
\begin{equation}
a_{i} = 1.5 \times 10^4 (2-3e_{\rm pl}) m_{\rm pl}^3,
\end{equation}
\begin{equation}
b_{i} = 1+\frac{m_{\rm pl}e_{\rm pl}^2}{10},
\end{equation}
\begin{equation}
c_{i} = \frac{1.2 \times 10^6}{\left(2-3e_{\rm pl}\right)\left(5+e_{\rm pl}^2\left[m_{\rm pl}+2\right]^3\right)},
\end{equation}
\begin{equation}
d_{i} = -3 + 2 e_{\rm pl},
\end{equation}
and
\begin{equation}
g_{i} = \sqrt{\frac{3 m_{\rm pl}}{e_{\rm pl}+0.001}}.
\end{equation}
To include the damping in the code, we converted it into an acceleration according to the formulas \citep{Papaloizou_2000}
\begin{equation}
\mathbf{a}_{\rm ecc} = -2\frac{(\mathbf{v}_{\rm pl} \cdotp \mathbf{r}_{\rm pl}) \mathbf{r}_{\rm pl}}{\left\| \mathbf{r}_{\rm pl}\right\|^2 \tau_{\rm ecc}},
\end{equation}
\begin{equation}
\mathbf{a}_{\rm inc} = -2 \frac{(\mathbf{v}_{\rm pl} \cdotp \mathbf{k}) \mathbf{k}}{\tau_{\rm inc}},
\end{equation}
where $\mathbf{v}_{\rm pl}$ and $\mathbf{r}_{\rm pl}$ are the velocity and the position of the planet, respectively, $\mathbf{k}$ is the unitary vertical vector,
\begin{equation}
\tau_{\rm ecc} = \left|\frac{e_{\rm pl}}{de/dt}\right|,
\end{equation}
and
\begin{equation}
\tau_{\rm inc} = \left|\frac{i_{\rm pl}}{di/dt}\right|.
\end{equation}
In Eq.~\eqref{e_dot}, the time is expressed in orbital period and the inclination damping in degrees per orbit. We refer to \cite{Sotiriadis_2017} for more details.


The code also includes the disk gravitational potential acting on the planets expressed in spherical coordinates as \citep{Terquem_2010,Hure_2011}
\begin{equation}
\Phi(r_{\rm pl},\varphi_{\rm pl},\theta_{\rm pl}) = -2G \int_{R_{\rm in}}^{R_{\rm out}} \Sigma(\tilde{r}) \sqrt{\frac{\tilde{r}}{r_{\rm pl} \sin(\theta_{\rm pl})}} k K(k) d\tilde{r},
\end{equation}
where, $(r_{\rm pl},\varphi_{\rm pl},\theta_{\rm pl})$ is the position of the planet in spherical coordinates and $K(k)$ is the complete elliptic integral of first kind:
\begin{equation}
K(k)=\int_{0}^{\pi/2} \frac{d\tilde{\varphi}}{\sqrt{1-k^2\sin^2\left(\tilde{\varphi}\right)}}
\end{equation}
with $k$ between 0 and 1 and equal, in our case, to 
\begin{equation}
k=2\sqrt{\frac{r_{\rm pl} \tilde{r} \sin(\theta_{\rm pl})}{r_{\rm pl}^2+\tilde{r}^2+2r_{\rm pl}\tilde{r}\sin(\theta_{\rm pl})}}.
\end{equation} 
We converted this potential in terms of acceleration in cartesian coordinates through \citep{Roisin_2021b} 
\begin{equation}
\begin{aligned}
\frac{d^2x_{\rm pl}}{dt^2} &= - \sin(\theta_{\rm pl}) \cos(\varphi_{\rm pl}) \frac{\partial \Phi}{\partial r_{\rm pl}} - \cos(\theta_{\rm pl}) \cos(\varphi_{\rm pl}) \frac{1}{r_{\rm pl}} \frac{\partial \Phi }{\partial \theta_{\rm pl}} \\
\frac{d^2y_{\rm pl}}{dt^2} &= - \sin(\theta_{\rm pl}) \sin(\varphi_{\rm pl}) \frac{\partial \Phi}{\partial r_{\rm pl}} - \cos(\theta_{\rm pl}) \sin(\varphi_{\rm pl}) \frac{1}{r_{\rm pl}} \frac{\partial \Phi }{\partial \theta_{\rm pl}} \\
\frac{d^2z_{\rm pl}}{dt^2} &= - \cos(\theta_{\rm pl}) \frac{\partial \Phi}{\partial r_{\rm pl}} + \sin(\theta_{\rm pl})  \frac{1}{r_{\rm pl}} \frac{\partial \Phi }{\partial \theta_{\rm pl}}, 
\end{aligned}
\end{equation}
where $(x_{\rm pl},y_{\rm pl},z_{\rm pl})$ is the cartesian coordinate of the planet and where
\begin{equation}
\begin{aligned}
\frac{\partial \Phi}{\partial r_{\rm pl}} &= \frac{G}{r_{\rm pl}} \int_{R_{\rm in}}^{R_{\rm out}} \Sigma(\tilde{r}) \sqrt{\frac{\tilde{r}}{r_{\rm pl} \sin(\theta_{\rm pl})}} k \left[K(k) - \frac{\tilde{r}^2-r_{\rm pl}^2}{R^2}E(k)\right] d\tilde{r} \\
\frac{\partial \Phi}{\partial \theta_{\rm pl}} &= G \int_{R_{\rm in}}^{R_{\rm out}} \Sigma(\tilde{r}) \cot(\theta_{\rm pl})\sqrt{\frac{\tilde{r}}{r_{\rm pl} \sin(\theta_{\rm pl})}} k  \Bigg[ K(k)  \\
&\hspace{5.5cm} \left.- \frac{\tilde{r}^2+r_{\rm pl}^2}{R^2}E(k)\right] d\tilde{r}
\end{aligned}
\label{derivative}
\end{equation}
with $R^2=r_{\rm pl}^2+\tilde{r}^2-2r_{\rm pl}\tilde{r}\sin(\theta_{\rm pl})$ and $E(k)$ the complete elliptic integral of the second kind: 
\begin{equation}
E(k)=\int_{0}^{\pi/2} \sqrt{1-k^2\sin^2\left(\tilde{\varphi}\right)} d\tilde{\varphi}.
\end{equation}

Note that all the accelerations presented in this section have been added in the code preserving the symmetry of SyMBA as prescribed in \cite{Lee_2002} : 
\begin{equation}
f(t_{k+1}) = E_{\rm disk}\left(\frac{\tau}{2}\right) \exp\left(T \tau \right)E_{\rm disk}\left(\frac{\tau}{2}\right) f(t_{k}),
\end{equation}
where $T$ is the linear operator associated with the Hamiltonian  $\mathcal{H}$ of the $N$-body system without the disk and $E_{\rm disk}\left(\cdot\right)$ is the evolution due to the disk.

Moreover, we considered the nodal precession of the disk induced by the binary companion as in \cite{Roisin_2021b}, which leads to a uniform precession of the disk's ascending node and a variation of the disk inclination with amplitude $2i_B$ with respect to the initial plane of the disk: 
\begin{equation}
\left\{
\begin{aligned}
&i_d(t) = 2 i_B\left|\sin(2\pi f_d t)\right| \\
&\Omega_d(t) = \frac{\pi}{2} + \Omega_B - \left( 2 \pi f_d t \mod \pi\right)
\end{aligned}\right. ,
\label{precess}
\end{equation}
where
$i_d$ and $\Omega_d$ are the inclination and longitude of the ascending node of the disk, $i_B$ and $\Omega_B$ are the inclination and longitude of the ascending node of the binary companion in the initial plane of the disk, and
\begin{equation}
f_d = 0.2145 \left(\frac{a_{\rm out}}{1 \rm au}\right)^{3/2} \left(\frac{a_B}{1 \rm au}\right)^{-3} \left(\frac{m_B}{1 m_\odot}\right),
\label{freq_disk}
\end{equation}
with $a_B$ and $m_B$ the semi-major axis and mass of the binary companion, respectively.

In our simulations, the disk mass decreases exponentially with time. We consider the disk as being fully dissipated and thus we neglect its influence when 
\begin{equation}
\frac{d{m}_{\rm disk}(t)}{dt}=\frac{m_{\rm disk,0}}{T_0} \exp{(-t/T_0})<10^{-9} m_0/yr,
\end{equation}
where $m_0$ denotes the mass of the central star, $m_{\rm disk,0}$ the disk mass at $t=0$ (fixed here to 8 $m_{\rm Jup}$), and $T_0=2.8\times 10^5$ yr. More information about the code can also be found in \cite{Roisin_2021b}.

As commonly adopted, we applied an inward migration to the outer planet only (this approach favors convergent migration), while the eccentricity and inclination damping due to the disk is applied on both planets. For the integration, we used a time step of $0.005$ yr for the disk phase and $0.01$ yr for the dynamical evolution after the dispersal of the disk. In both cases, a recursive reduction of the timestep is implemented in case of close encounters, as described in \cite{Duncan_1998}. The simulations were carried out for 100 Myr. In the following, subscripts $1$ and $2$ will refer to the inner planet and outer planet, respectively.

\begin{figure*}
	\centering
	\includegraphics[width=0.9\columnwidth]{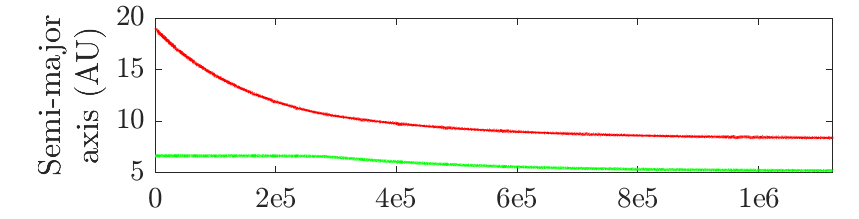}
	\includegraphics[width=0.9\columnwidth]{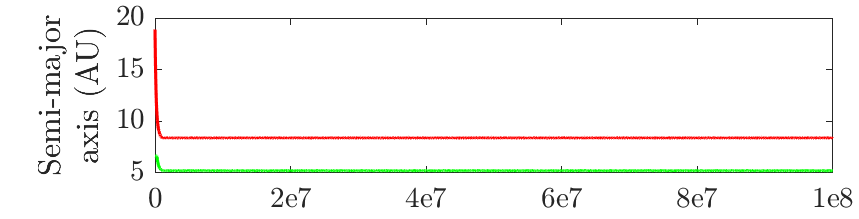}
	
	\includegraphics[width=0.9\columnwidth]{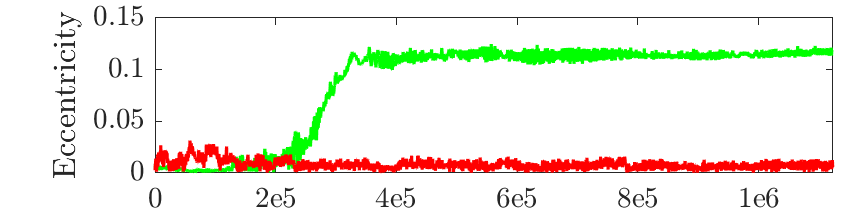}
	\includegraphics[width=0.9\columnwidth]{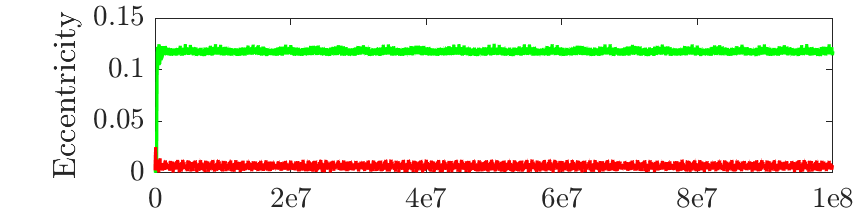}
	
	\includegraphics[width=0.9\columnwidth]{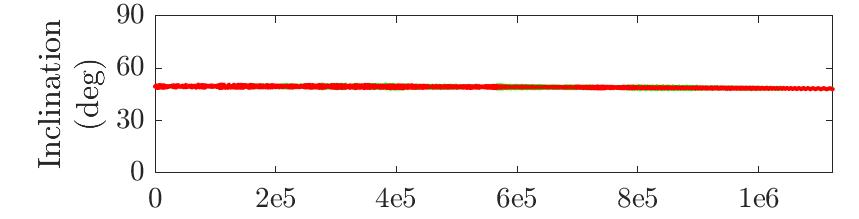}
	\includegraphics[width=0.9\columnwidth]{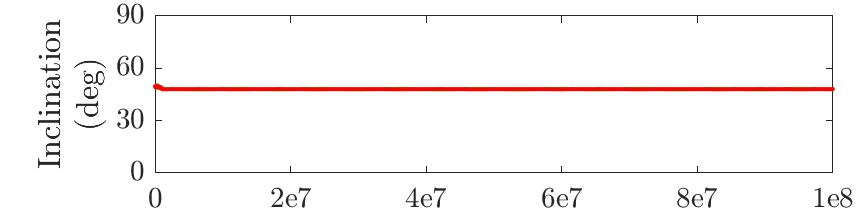}
	
	\includegraphics[width=0.9\columnwidth]{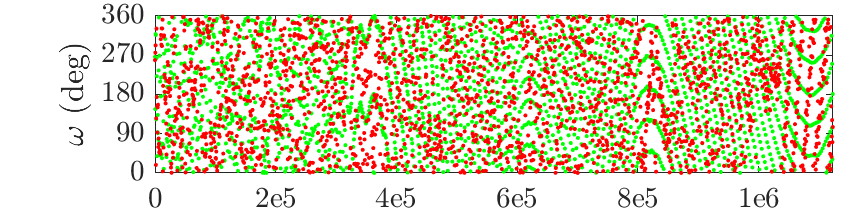}	\includegraphics[width=0.9\columnwidth]{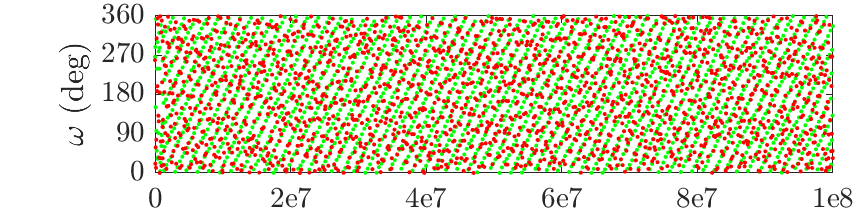}
	
	\includegraphics[width=0.9\columnwidth]{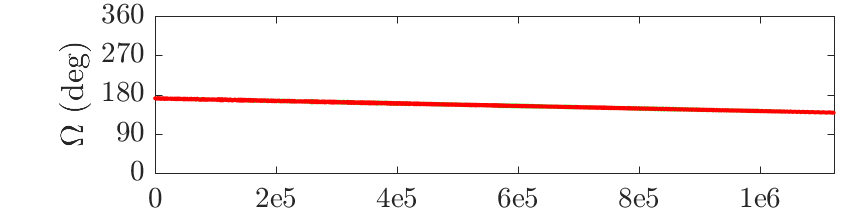}
	\includegraphics[width=0.9\columnwidth]{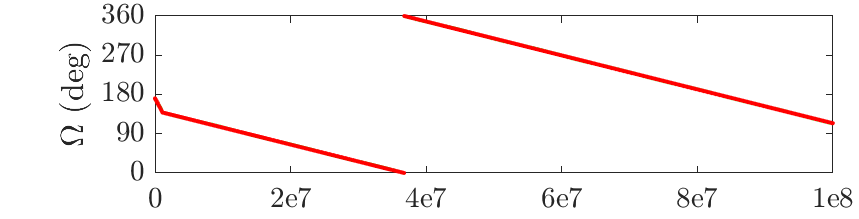}
	
	\includegraphics[width=0.9\columnwidth]{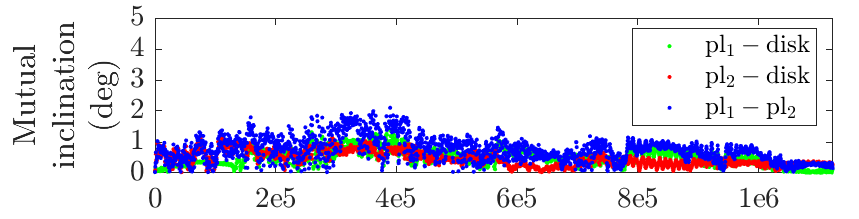}
	\includegraphics[width=0.9\columnwidth]{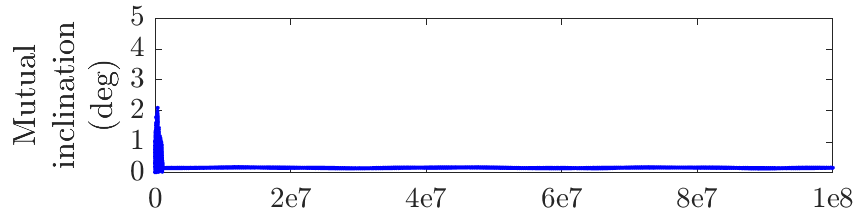}
	
	\includegraphics[width=0.9\columnwidth]{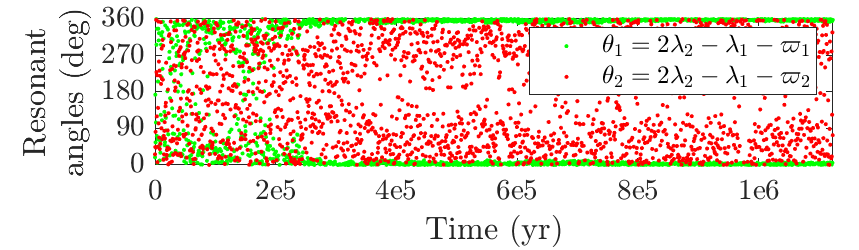}
	\includegraphics[width=0.9\columnwidth]{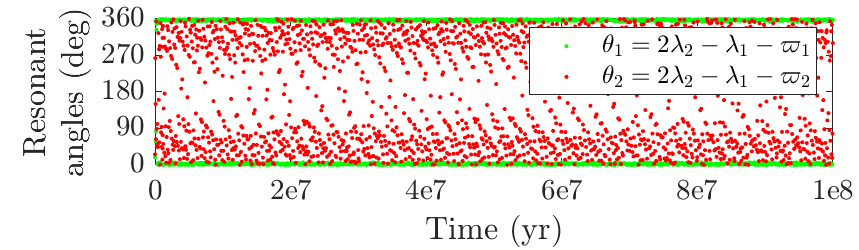}
	\caption{Typical evolution of a system of two giant planets with a wide binary at $a_B=1000$ au (in the Laplace plane reference frame), during the disk phase (left panels) and after the disk dispersal (right panels). The initial parameters (with respect to the disk plane) are $e_B=10^{-3}$ and $i_B=50^\circ$ for the binary companion and $a_{1}=6.7$ au, $m_{1}=1.44~M_{\rm Jup}$, $a_{2}=18.9$ au, and $m_{2}=3.55~M_{\rm Jup}$ for the planets. }
	\label{typ_disk1}
\end{figure*}

\begin{figure*}
	\centering
	\includegraphics[width=0.9\columnwidth]{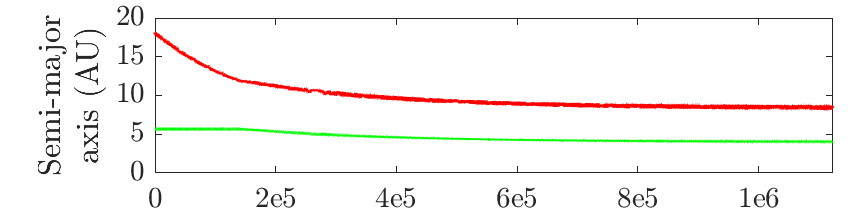}
	\includegraphics[width=0.9\columnwidth]{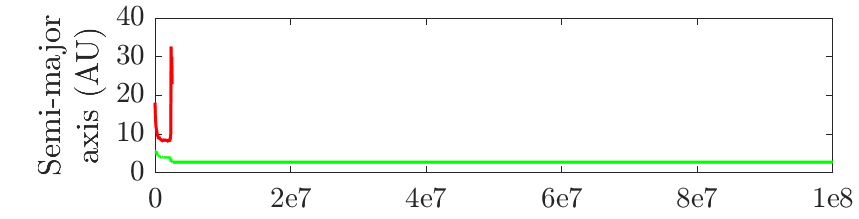}
	
	\includegraphics[width=0.9\columnwidth]{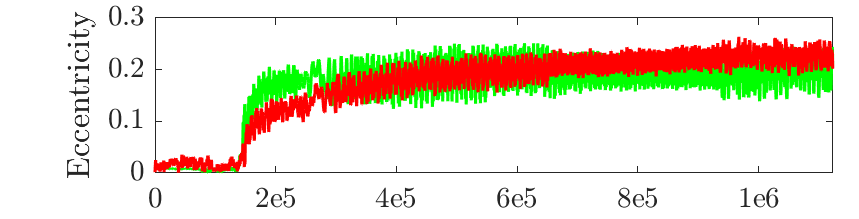}
	\includegraphics[width=0.9\columnwidth]{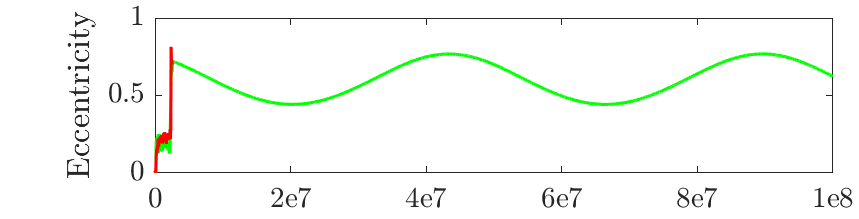}
	
	\includegraphics[width=0.9\columnwidth]{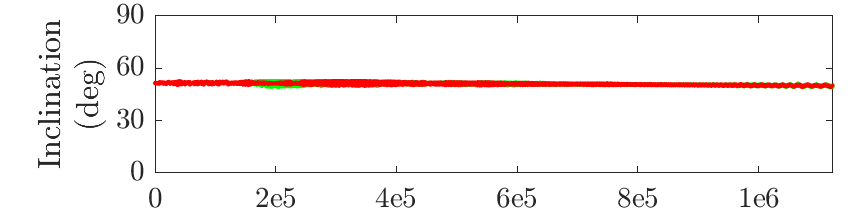}
	\includegraphics[width=0.9\columnwidth]{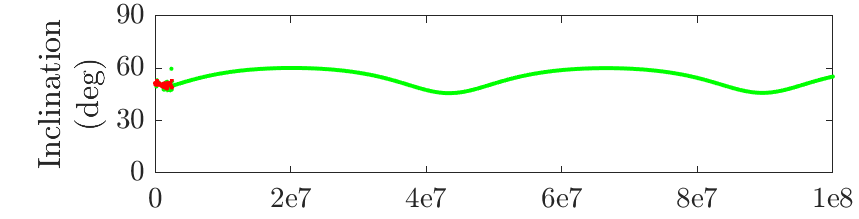}
	
	\includegraphics[width=0.9\columnwidth]{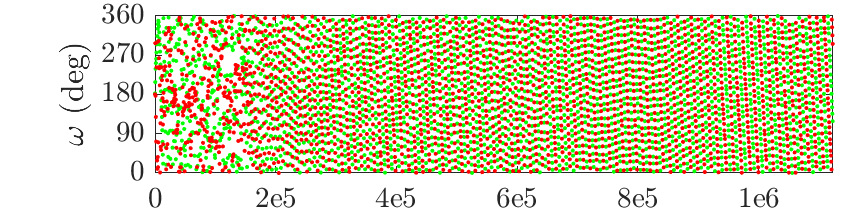}	\includegraphics[width=0.9\columnwidth]{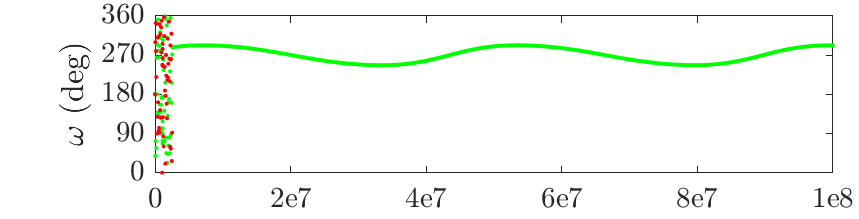}
	
	\includegraphics[width=0.9\columnwidth]{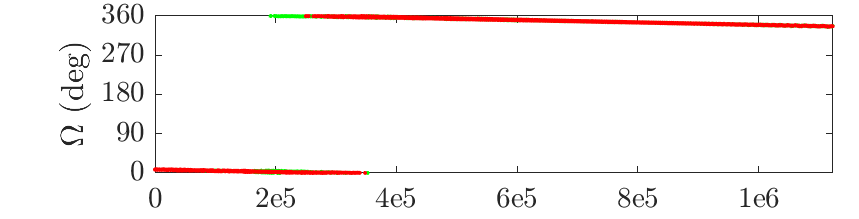}
	\includegraphics[width=0.9\columnwidth]{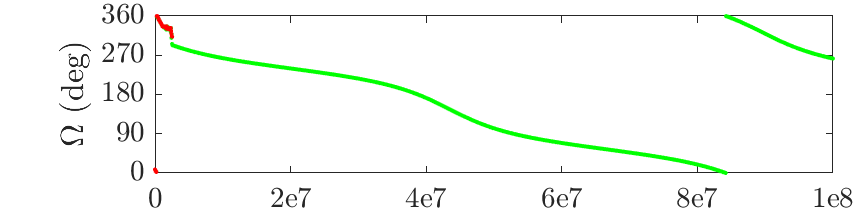}
	
	\includegraphics[width=0.9\columnwidth]{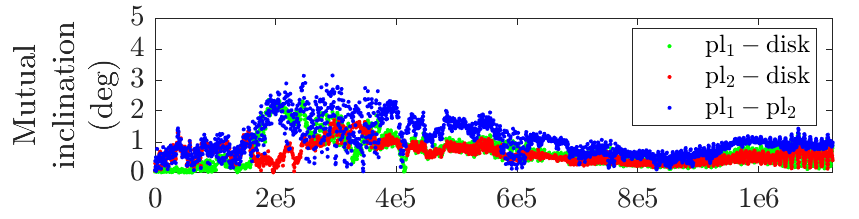}
	\includegraphics[width=0.9\columnwidth]{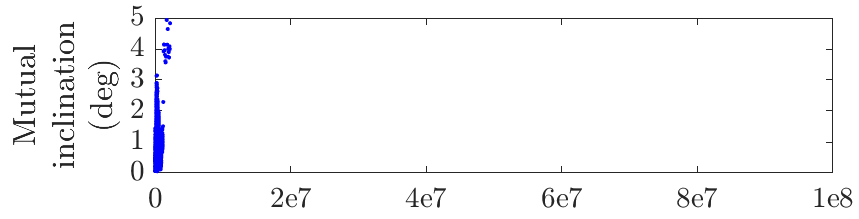}
	
	\includegraphics[width=0.9\columnwidth]{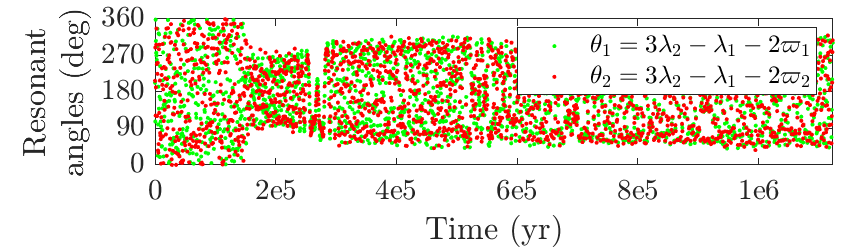}
	\includegraphics[width=0.9\columnwidth]{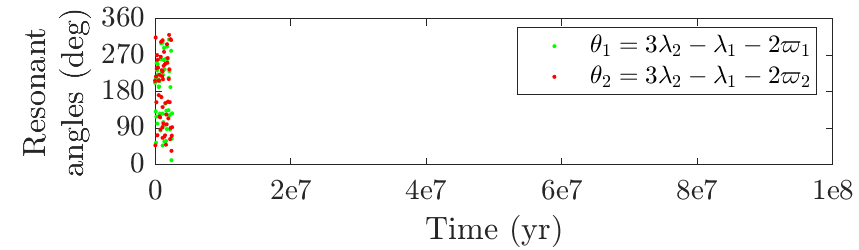}
	\caption{Same as Fig.~\ref{typ_disk1}, for the initial parameters $e_B=0.5$, $i_B=50^\circ$, $a_{1}=5.7$ au, $m_{1}=2.77~M_{\rm Jup}$, $a_{2}=18.15$ au, and $m_{2}=2.99~M_{\rm Jup}$. }
	\label{typ_disk2}
\end{figure*}

\section{Timescale analysis}
\label{sec:timescales}

There are several timescales relevant to our study: the disk's lifetime, the timescale of the planet-planet interactions, and the timescale of the precession induced by the binary  (i.e., the ZLK resonance timescale). Since we will use them to interpret the outcomes of our simulations throughout the paper, we review them first.

All effects inducing a precession of the pericenter of a planet faster than the one induced by the ZLK resonance act to suppress it. It is the case with, e.g., general relativity, tidal and rotational bulges and, as is the case here, gravitational interaction with another planet \citep{Wu_2003,Fabrycky_2007,Takeda_2008}.
The timescale associated with the ZLK resonance induced by a binary companion, $\tau_{\rm ZLK}$, can be approximated by \citep{Kiseleva_1998}
\begin{equation}
\tau_{\rm ZLK} = \frac{2}{3\pi} \frac{P_B^2}{P_j}(1-e_B^2)^{3/2} \frac{m_0+m_j+m_B}{m_B},
\label{tau_LK}
\end{equation}
where $P_B$ and $P_j$ are the orbital periods of the binary companion and the $j$-th planet, respectively, and $m_j$ the mass of the $j$-th planet ($j=1,2$). If the ZLK timescale if longer than the disk's lifetime, then the binary companion does not significantly perturb the orbits of the planets while they migrate in the disk.

The timescale of the pericenter precession due to the gravitational interaction between two planets, denoted $\tau_{\rm pl}$, can be estimated by using the Laplace-Lagrange secular theory. It can be approximated by means of the eigenvalues of the following matrix \citep{Brouwer_1961,Murray_1999,Takeda_2008}:
\begin{equation}
A=
\begin{pmatrix}
c_1 & -c_0 c_1 \\
-c_0 c_2 & c_2
\end{pmatrix}
\end{equation}
with 
\begin{equation}
\begin{aligned}
c_0&=\frac{b_{3/2}^{(2)}\left(\alpha\right)}{b_{3/2}^{(1)}\left(\alpha\right)} \\
c_1&=\frac{1}{4}n_1\frac{m_2}{m_0+m_1} \alpha^2b_{3/2}^{(1)}\left(\alpha\right) \\
c_2&=\frac{1}{4}n_2\frac{m_1}{m_0+m_2} \alpha^2b_{3/2}^{(1)}\left(\alpha\right),
\end{aligned}
\end{equation}
where $b_{3/2}^{(1)}(\alpha)$ and $b_{3/2}^{(2)}(\alpha)$ are the Laplace coefficients of the first and second kind, respectively, $\alpha=a_1/a_2$ the planetary semi-major axis ratio, and $n_j$ the planetary mean motions ($j=1,2$). The expressions of the eigenvalues are \citep{Zhou_2003}: 
\begin{align}
g_+&=\frac{1}{2} \left(c_1+c_2+\sqrt{(c_1-c_2)^2+4c_0^2c_1c_2}\right) \label{eig_+}\\
g_-&=\frac{1}{2} \left(c_1+c_2-\sqrt{(c_1-c_2)^2+4c_0^2c_1c_2}\right). \label{eig_-}
\end{align}
The timescale of the pericenter precession $\tau_{\rm pl}$ for each planet is thus given by $2\pi/g_+$ and $2\pi/g_-$ with the highest value associated with the more massive planet. This approximation is accurate under the condition presented in \cite{Takeda_2008}, namely 
\begin{equation}
\frac{\alpha}{1-3q\sqrt{\alpha}/b_{3/2}^{(1)}(\alpha)} \ll 1
\end{equation}
with $q=m_2/m_1$. In our work, the condition is not always fulfilled but even when it is not the case, the formula will still give us an indication about the order of magnitude of the precession timescale.

\section{Wide binary with $a_B=1000$ au}
\label{sec:wide_bin}

\begin{table}
	\caption{Initial parameters for the simulations. The orbital elements are expressed with respect to the initial plane of the disk.}
	\centering
	\begin{tabular}{lrrrr}
		\hline
		& Inner planet & Outer planet & Binary companion\\
		\hline
		mass  & $U[1;5]$ $M_{\rm Jup}$ & $U[1;5]$ $M_{\rm Jup}$ & $1$ $M_{\odot}$\\
		$a$ (au)  & $U[5,7]$ & $U[12,21]$ & $1000$\\
		$e$  & $U[10^{-3};0.01]$ & $U[10^{-3};0.01]$ & $10^{-3}$, $0.1$, $0.3$, $0.5$\\ 
		$i$ ($^\circ$) & $U[0.01;0.1]$ & $U[0.01;0.1]$ & $10^{-3}$, $10$, $20$, $30$,   \\
		& & & $40$,$50$, $60$, $70$ \\
		$\Omega$ ($^\circ$) & $U[0,360]$ &  $U[0,360]$& $U[0,360]$\\
		$\omega$ ($^\circ$) & $U[0,360]$ &  $U[0,360]$& $U[0,360]$\\
		$M$ ($^\circ$) & $U[0,360]$ &  $U[0,360]$& $U[0,360]$\\
		\hline
	\end{tabular}
	
	\label{Body_param2}
\end{table}

In this section we start by considering a binary companion at $a_B=1000$~au. At such a separation, the timescale of the perturbations arising from the binary is longer than the disk's lifetime, and the binary will have a minimal impact on the migrating planets. We use this setup to smoothly form planets in mean-motion resonances (MMRs) via disk migration. We then evolve the simulations for another 100~Myr, in order to study the long-term impact of the binary on the planets, with emphasis on the resonant dynamics. In the next section, closer binaries will be considered, which will have a visible impact on the planets as early as in the migration phase.

Regarding the initial parameters of the simulations, the masses and orbital elements of the two giant planets and the binary companion are given in Table~\ref{Body_param2}. For the binary, eight different inclinations (with respect to the initial plane of the disk) ranging from nearly $0^\circ$ to $70^\circ$, as well as four different eccentricities ($10^{-3}$, 0.1, 0.3, and 0.5) are considered here. The semi-major axes of the planets are chosen randomly between 5 and 7 au for the inner planet and between 12 and 21 au for the outer planet, in order to allow captures in different MMRs during the migration. The closest initial location of the two planets is beyond the 2:1 MMR ($(12/7)^{3/2}\simeq 2.24$), while the furthest one is beyond the 8:1 MMR ($(21/5)^{3/2}\simeq 8.61$). The planetary masses, eccentricities, and inclinations are chosen randomly as $m\sim U [1;5] m_{\rm Jup}$, $e \sim U [0.001; 0.01]$, and $i \sim U [0.01^\circ;0.1^\circ]$. For each combination of the fixed parameters of the binary companion, we drew 100 different random initial parameters for the planets.

\begin{figure}[]
	\centering
	\includegraphics[width=0.98\columnwidth]{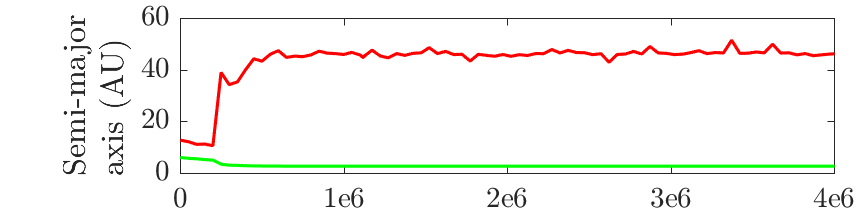}
	\includegraphics[width=0.98\columnwidth]{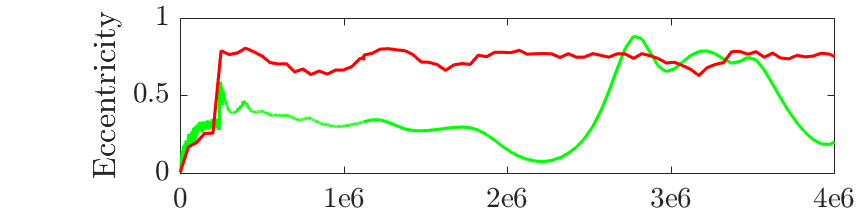}
	\includegraphics[width=0.98\columnwidth]{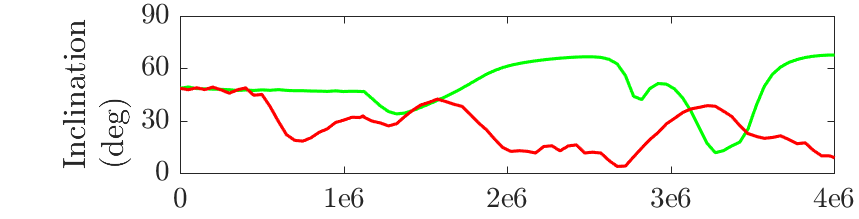}
	\includegraphics[width=0.98\columnwidth]{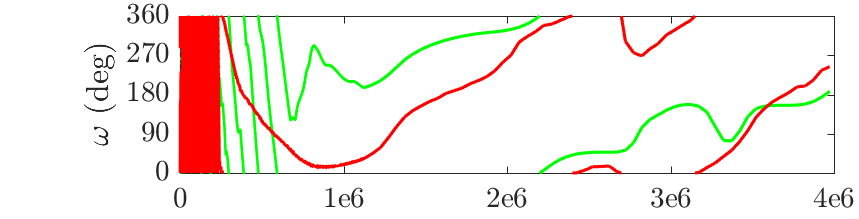}
	\includegraphics[width=0.98\columnwidth]{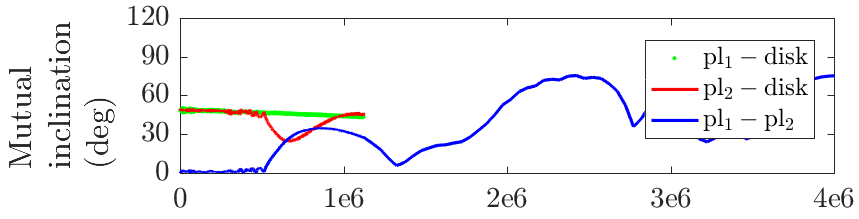}
	\includegraphics[width=0.98\columnwidth]{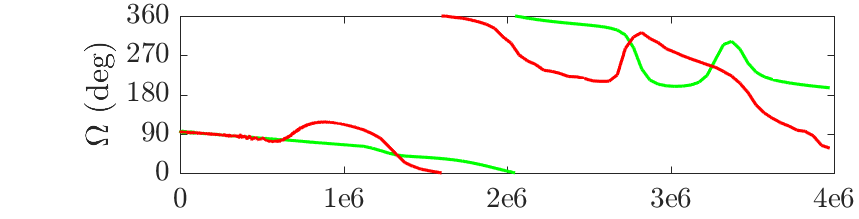}
	\includegraphics[width=0.98\columnwidth]{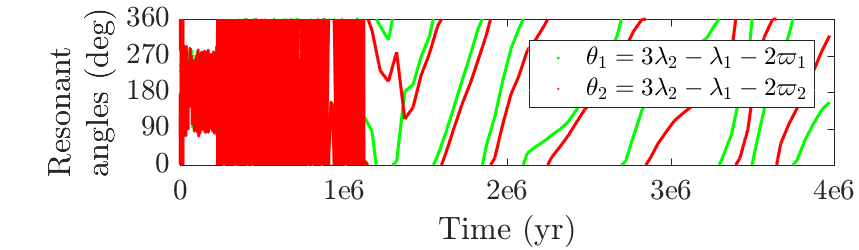}
	\caption{Same as Fig.~\ref{typ_disk1}, for the initial parameters $e_B=0.1$, $i_B=50^\circ$, $a_{1}=6$ au, $m_{1}=4.11~M_{\rm Jup}$, $a_{2}=12.74$ au, and $m_{2}=3.47~M_{\rm Jup}$.}
	\label{typ_disk3}
\end{figure}

\subsection{Typical examples}
\label{typ_exa}
We start by presenting several typical evolutions observed in the simulations. For each case, we will discuss both the disk phase and the long-term dynamical evolution after the dispersal of the disk. The evolutions are shown in the Laplace plane reference frame which is more suitable for the study of the ZLK resonance since the latter can easily be identified in this reference frame from the libration of the planetary pericenter argument \citep[e.g.,][]{Libert_2009}. The Laplace plane is the invariant plane orthogonal to the total angular momentum of the system that almost corresponds, in our case, to the plane of the binary star. In this particular reference plane, the disk does not change in inclination but only precesses as shown in \cite{Roisin_2021b} (see their section 2.3.2).

A first example is shown in Fig.~\ref{typ_disk1}, for a binary companion with $e_B=10^{-3}$ and $i_B=50^\circ$ and planetary masses of $m_{1}=1.44~M_{\rm Jup}$ and $m_{2}=3.55~M_{\rm Jup}$. The evolution during the disk phase is shown in the left panels. We observe that the outer planet migrates inwards and is captured at about 0.25 Myr in the 2:1 MMR with the inner planet, as confirmed by the libration of both resonant angles $\theta_1=2\lambda_2-\lambda_1-\varpi_1$ and $\theta_2=2\lambda_2-\lambda_1-\varpi_2$ (last panel), with $\lambda=\omega+\Omega+M$ and $\varpi=\omega+\Omega$ being the mean longitude and the longitude of the pericenter, respectively. After the capture in MMR, the two planets start migrating together and the eccentricity of the inner planet increases while the outer planet is kept on a quasi circular orbit \citep[see, e.g.,]{Lee_2002}. The eccentricity increase is limited due to the eccentricity damping by the disk and is followed by a slight increase of the mutual inclination between the planets which is rapidly damped by the disk. The disk keeps the planets within its midplane during the whole disk phase (see the mutual inclination between the disk and the planets in the sixth left panel), despite the influence of the inclined binary companion. Regarding the dynamical evolution of the system after the dispersal of the disk (right panels of Fig.~\ref{typ_disk1}), we observe that the MMR is preserved during the whole simulation. We also notice the linear evolutions of the planetary longitudes of the ascending nodes which, coupled with the constant inclinations, indicate the nodal precession of the planets due to the binary companion. Moreover, there is no evidence of a ZLK resonance with the highly inclined binary companion (i.e., no libration of the pericenter arguments around $90^\circ$ or $270^\circ$). This is due to the strong gravitational interaction between the planets.


A second system is displayed in Fig.~\ref{typ_disk2}. Its evolution is similar to the one of the previous system during the disk phase, except now the planets get captured in the 3:1 MMR instead (left panels). However, after the disk phase (right panels), a scattering event rapidly occurs leading to the ejection of the outer planet. The remaining planet is captured in a ZLK resonance with the binary companion, as shown by the libration of the pericenter argument around $270^\circ$ in the fourth right panel of Fig.~\ref{typ_disk2} as well as the coupled variation in the planetary eccentricity and inclination.

In some cases, a destabilization of the system takes place during the disk phase, as illustrated in Fig.~\ref{typ_disk3}. At about 0.25~Myr, while the planets evolve in the 3:1 MMR and pursue their migration, a scattering event occurs. The outer planet is scattered to a larger separation and does not evolve in the disk anymore. The evolution of both planets rapidly becomes chaotic. They alternate between ZLK resonant states and non-resonant evolutions (see the evolution of the planetary arguments of pericenters), until the disruption of the system at about 6~Myr. The instability comes from the nearly equal strength of the ZLK effect induced by the binary and the gravitational interaction between the planets, as will be shown in Sect.~\ref{sec:close_bin}.
\begin{figure}[]
	\centering
	\includegraphics[width=0.9\columnwidth]{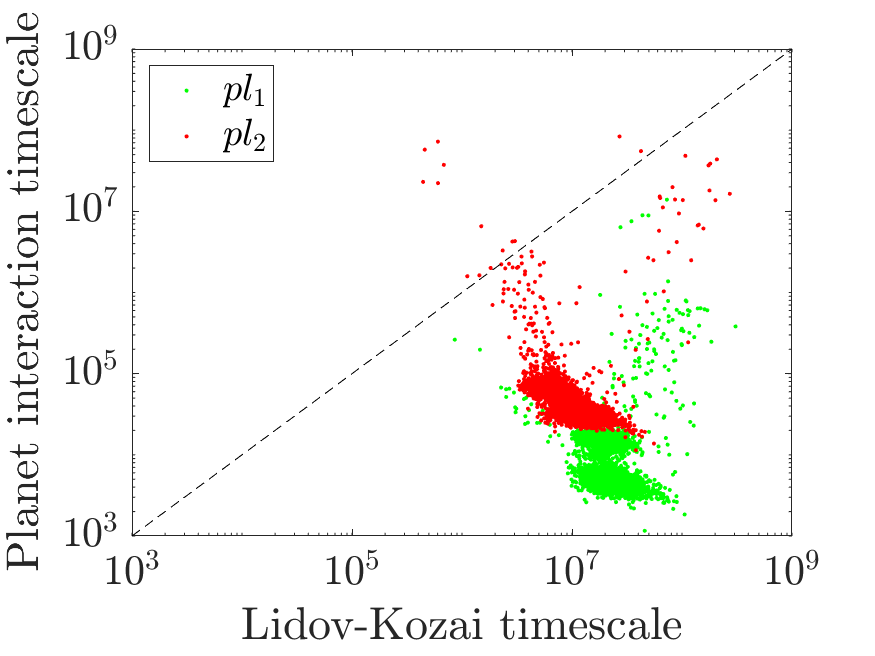}
	\caption{Timescale of the pericenter precession due to the gravitational interaction between the planets versus the ZLK timescale due to the binary companion. The green and red colors refer to the inner planet and outer planet, respectively.}
	\label{freq_comp}
\end{figure}
\begin{figure}[]
	\centering
	\includegraphics[width=0.49\columnwidth]{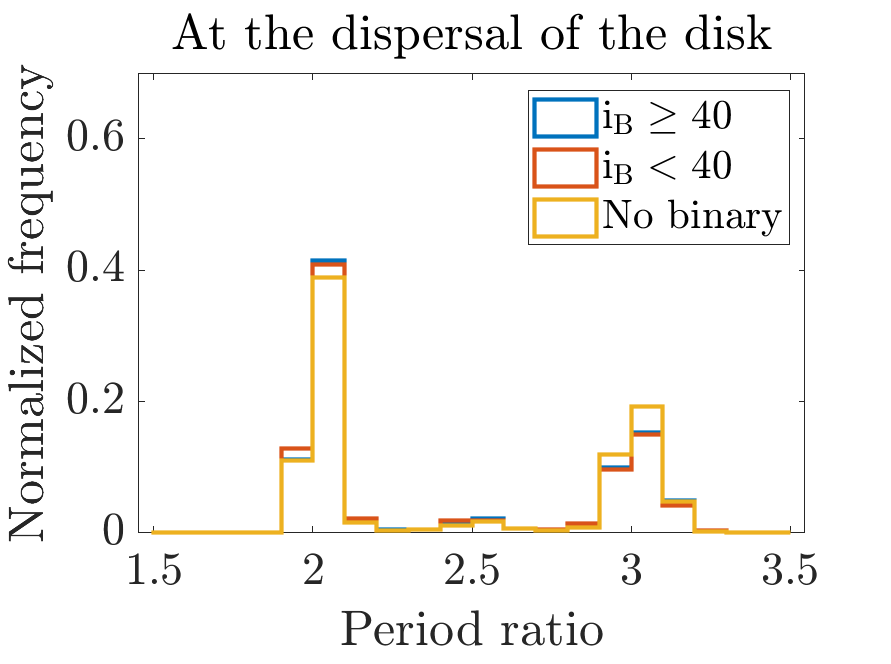}
	\includegraphics[width=0.49\columnwidth]{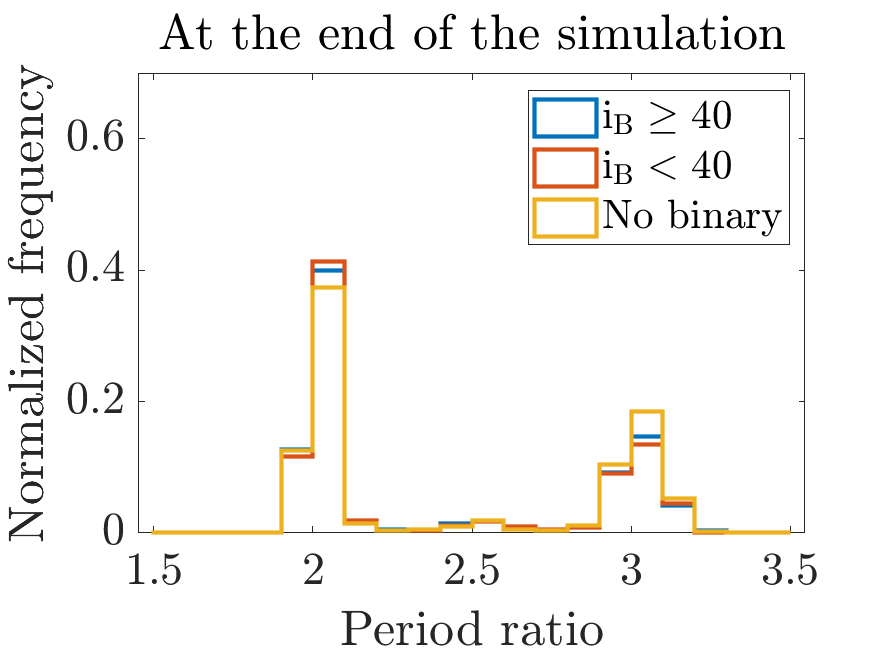}
	\includegraphics[width=0.49\columnwidth]{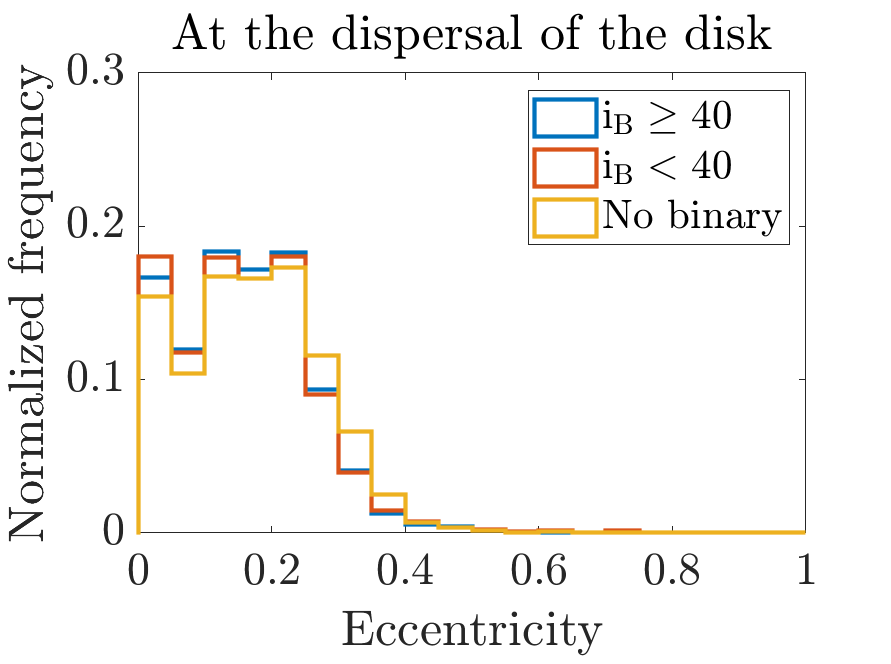}
	\includegraphics[width=0.49\columnwidth]{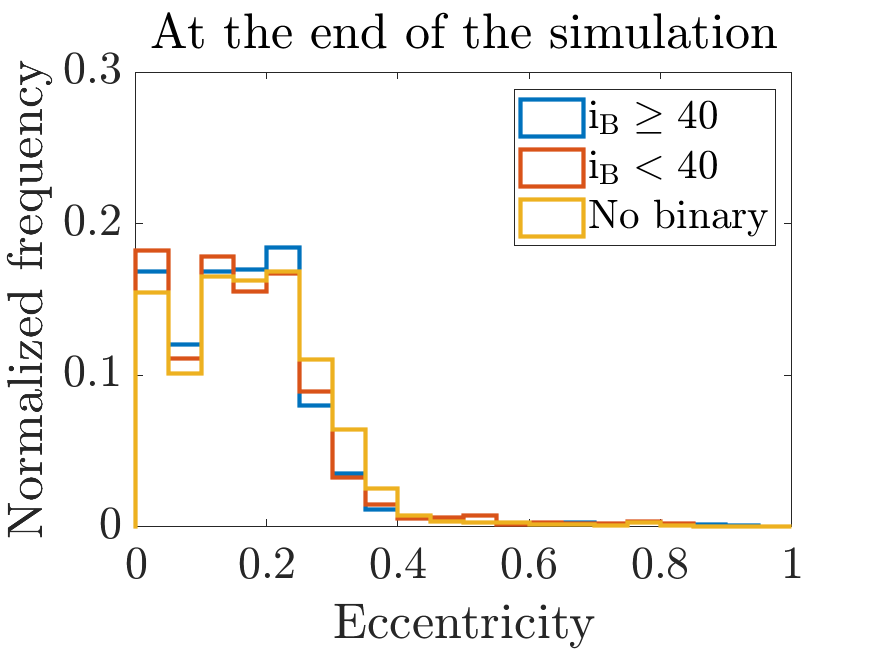}
	\includegraphics[width=0.49\columnwidth]{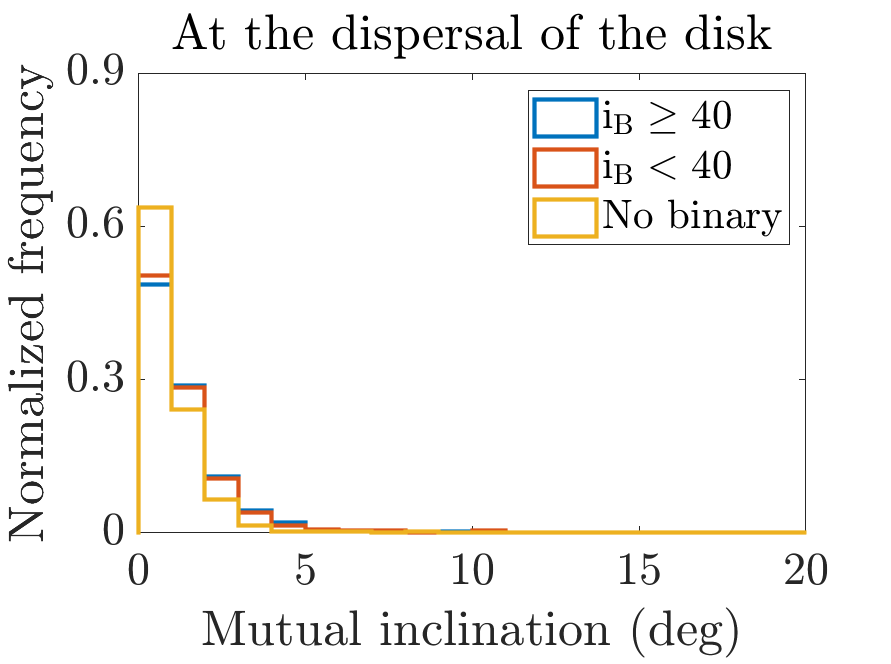}
	\includegraphics[width=0.49\columnwidth]{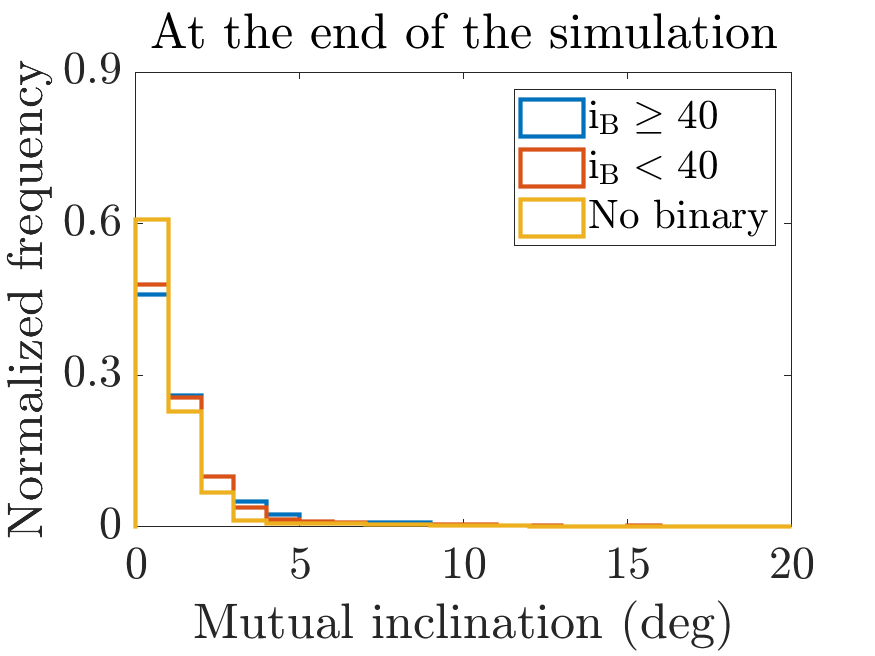}
	\includegraphics[width=0.49\columnwidth]{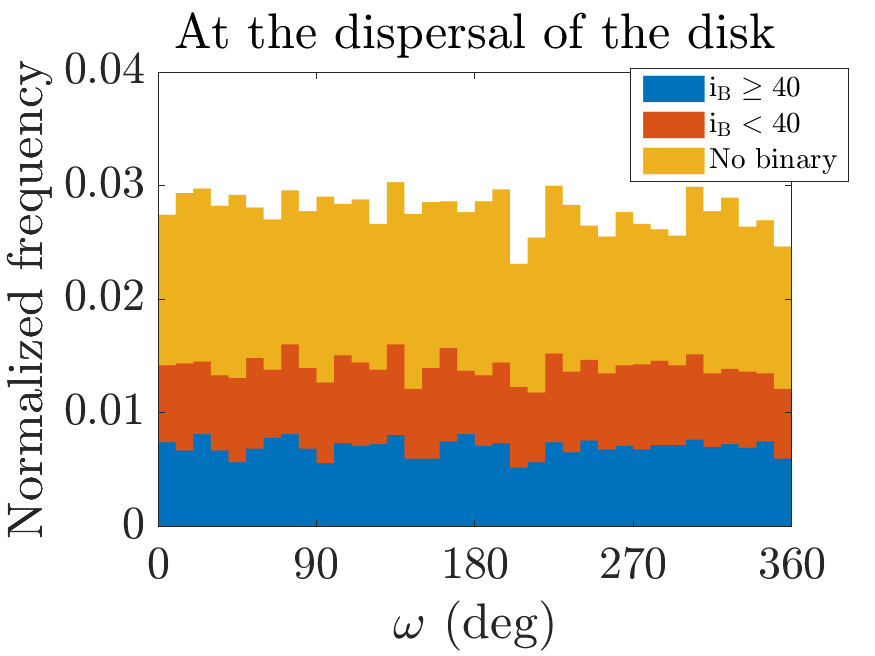}
	\includegraphics[width=0.49\columnwidth]{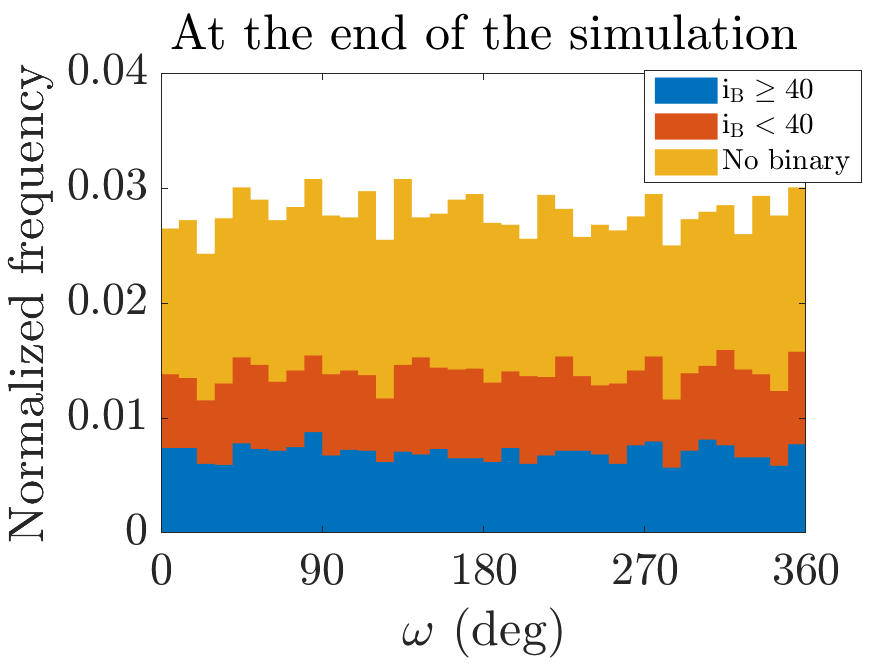}
	\caption{Normalized distributions of final orbital elements at the dispersal of the disk (left panel) and at the end of the simulation (right panel). From top to bottom: period ratio, eccentricity, mutual inclination, and argument of the pericenter (in the Laplace plane reference frame). The color code refers to the presence of a binary companion and its inclination value for the systems with a companion.}
	\label{P_ratio}
\end{figure}


\subsection{Results from a large ensemble of simulations}

We now study the system configurations found in the 3200 simulations performed with $a_B=1000$~au. In order to interpret these simulations, it is useful to consider the relevant timescales. The ZKL and planet-planet interaction timescales are shown in Fig.~\ref{freq_comp}, calculated with the system parameters found at the dispersal of the disk. As expected, for nearly all the systems, the ZLK timescale is much longer than the timescale associated with the gravitational interaction between the two planets, which explains that the influence of the binary is very limited in the simulations. In addition, the ZKL timescale is longer than the disk's lifetime, indicating that the binary had little influence over the planets during the migration process. As a result, here we focus on the long-term dynamics after the disk phase, which a particular emphasis on the interplay between planetary MMRs and binary perturbations.

The distributions of the final orbital parameters are displayed in Fig.~\ref{P_ratio}. To make it clearer, we present separately the systems for which the inclination of the binary companion is strictly below $40^\circ$ (orange color) and the other systems for which the binary companion is highly inclined ($i_B\geq40^\circ$) and which are thus likely to experience a ZLK resonance with the binary companion (blue color). Let us note that we also run the same 3200 simulations without the binary companion for comparison (yellow color). The distributions are shown at the dispersal of the disk in the left panels and at the end of the simulations in the right panels.


\subsubsection{MMR captures}

We first look at the period ratios between the planets (top panels of Fig.~\ref{P_ratio}) to study the MMR captures during the disk phase. Two resonant commensurabilities are more prevalent, namely the 2:1 MMR followed by the 3:1 MMR, although several systems are also trapped close to the 5:2 MMR. Overall, the effect of the wide binary companion is small, as expected from the timescale analysis. We see that the systems with a binary companion (blue and orange) are captured slightly more often in the 2:1 MMR and less captured in the 3:1 MMR than the systems without a binary companion (yellow curve). No significant influence of the inclination of the binary companion on the resonance capture is visible. The left and right panels look very similar, which leads us to the conclusion that the systems captured in a MMR during the disk phase stay trapped for a long time (100~Myr in this case), even under the influence of the binary companion. This shows that for the wide binary case investigated here, the chaotic evolutions after the disk phase are rather limited in two-planet resonant systems.

\begin{table}
	\centering
	\caption{Percentages of the different system configurations at the dissipation of the disk}
	\begin{tabular}{ccccc}
		\hline
		Final  & MMR & Binary  & Binary  & No  \\
		systems & &  ($i_B\geq40^\circ$) & ($i_B<40^\circ$) &binary \\
		\hline
		2 planets & 2:1 & 53.37 & 53.87 & 49.7\\
		&3:1 & 28.56 & 27.87 & 34.4\\
		&5:2 & 0.68 & 0.63 & 1.16 \\
		&other MMR & 0.15 & 0.25 & 1.58\\
		&no MMR & 14.5 & 14.88 & 10.66\\
		\hline
		1 planet& & 2.43 & 2.5 & 2.34 \\
		\hline
		no planet& & 0.31 & 0 & 0.16\\
		\hline
	\end{tabular} 
	\label{MMR_disk}
\end{table}
\begin{table}
	\centering
	\caption{Percentages of the different system configurations at the end of the simulation}
	\begin{tabular}{ccccc}
		\hline
		Final & MMR & Binary  & Binary  & No  \\
		systems & &  ($i_B\geq40^\circ$) & ($i_B<40^\circ$) & binary\\
		\hline
		2 planets &2:1 & 53.44 & 54.06 & 49.8\\
		&3:1 & 26.63 & 25 & 33.03\\
		&5:2 & 0.56 & 0.25 & 0.72\\
		&other MMR & 0.12 & 0.26 & 1.65\\
		&no MMR & 12.81 & 13 & 9.78 \\
		\hline
		1 planet& & 6.06 & 7.43 & 4.9 \\
		\hline
		no planet& & 0.38 & 0 & 0.16 \\
		\hline
	\end{tabular}  
	\label{MMR_end}
\end{table}

To confirm the previous results, we studied the evolution of the resonant angles for each system. In practice, the planets are considered locked in a MMR if at least one of the resonant angles librates with an amplitude of maximum $270^\circ$ during the 0.2 Myr before the time considered, namely the dissipation of the disk or the end of the simulation. The results regarding the MMR captures are gathered in Tables~\ref{MMR_disk} and \ref{MMR_end}. As previously observed, there are slightly more two-planet systems captured in the 2:1 MMR for systems with a binary companion (53.6\% against 49.7\% at the dispersal of the disk, see Table~\ref{MMR_disk}) while it is the contrary for the 3:1 MMR (28.2\% against 34.4\%) and the 5:2 MMR (0.66\% against 1.2\%). Let us note that roughly the same percentages of planetary ejections at the dispersal of the disk are observed for single star systems and binary star systems (see the two last rows of Table~\ref{MMR_disk}). In Table~\ref{MMR_end}, we see that the percentages of final one-planet systems have increased, in particular for binary star systems due to chaotic events occurring after the dispersal of the disk. Examples of chaotic evolutions after the disk phase were shown in Figs.~\ref{typ_disk2} and \ref{typ_disk3}. A study of the stability of the two-planet systems formed here is done in Section~\ref{sec:stability}. 



\begin{figure}[]
	\centering
	\includegraphics[width=.98\columnwidth]{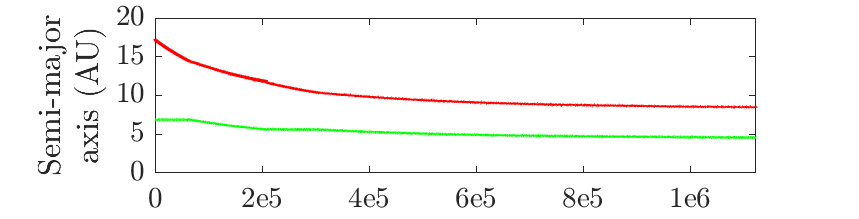}
	\includegraphics[width=.98\columnwidth]{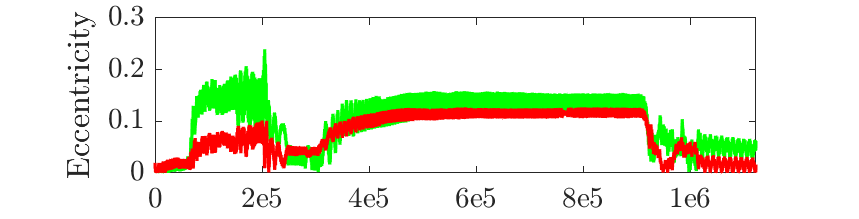}
	\includegraphics[width=.98\columnwidth]{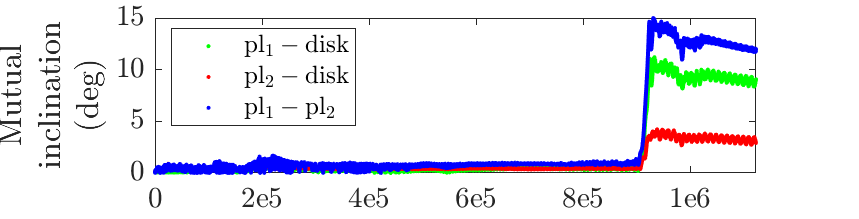}
	\includegraphics[width=.98\columnwidth]{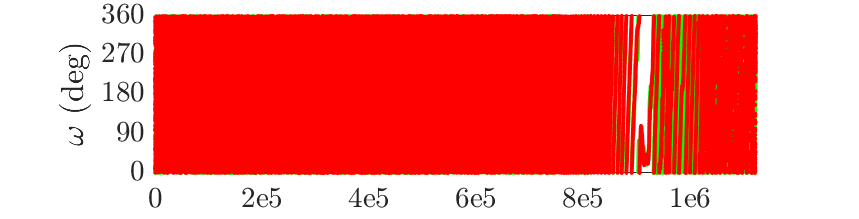}
	\includegraphics[width=.98\columnwidth]{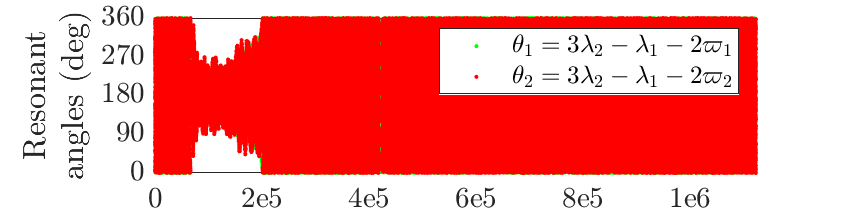}
	\includegraphics[width=.98\columnwidth]{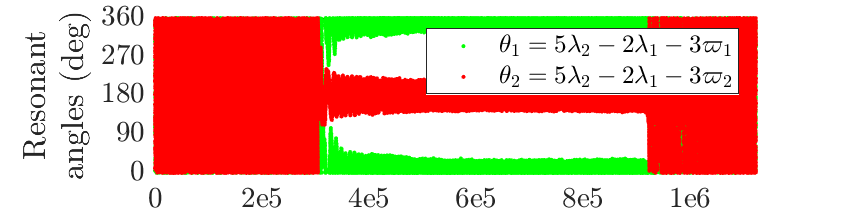}
	\includegraphics[width=.98\columnwidth]{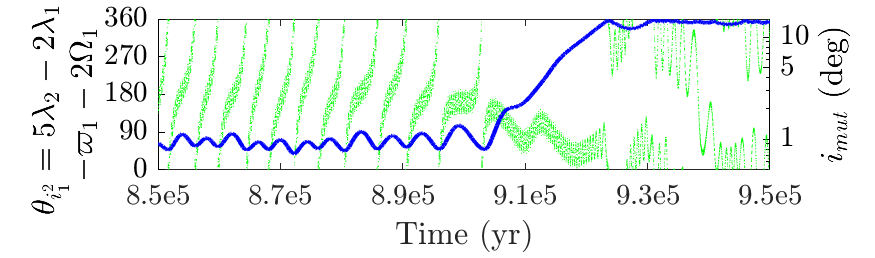}
	\caption{Typical evolution of two giant planets (in a wide binary star system) experiencing an inclination-type resonance during the disk phase. The initial parameters are $e_B=0.5$, $i_B=0.001^\circ$, $a_{1}=6.86$ au, $m_{1}=1.64~M_{\rm Jup}$, $a_{2}=17.1$ au, and $m_{2}=3.79~M_{\rm Jup}$.}
	\label{typ_disk4}
\end{figure}

\subsubsection{Eccentricities}

The eccentricity distribution of the different systems is shown in the panels of the second row of Fig.~\ref{P_ratio}. We observe that the single star systems with two planets have slightly higher eccentricities (orange and blue curves above the yellow curve for eccentricities below 0.25 while the yellow curve is above the two others for higher values). This observation is coherent with the percentages of resonance captures. The proportion of two-planet systems which are found in MMR is indeed more important for single star systems (see Tables~\ref{MMR_disk} and \ref{MMR_end}) and those resonance captures are generally followed by an increase of the planetary eccentricities. Again, the similar profile between the curves in the left panel (at the dispersal of the disk) and the right panel (at the end of the simulation) leads us to say that the wide binary companion has a limited influence on the system even after the dispersal of the disk.

\subsubsection{Inclinations}

In the third row panels of Fig.~\ref{P_ratio}, we show the distribution of the mutual inclination between the planets. We observe mutual inclination values slightly higher for systems including a binary companion. However, most of the mutual inclinations are below $5^\circ$. Only 3\% of the systems have a mutual inclination above $5^\circ$ in binary star systems. These higher mutual inclinations are generally produced by an inclination-type resonance. This mechanism is known to operate at large eccentricities in two-planet systems \citep{Thommes_2003,Libert_2009,Teyssandier_2014,Sotiriadis_2020}, but can also  be present at small to moderate eccentricities in systems with more than two planets \citep{Libert_2018}. An example of a highly mutually inclined system is shown in Fig.~\ref{typ_disk4}. We first observe a temporary capture in the 3:1 MMR, before a second capture in the 5:2 MMR. The eccentricities increase to moderate values and the inclinations suddenly grow at $9\times10^5$ yr when the system enters an inclination-type resonance. To better identify the resonant process leading to the inclination increase, we display at the bottom panel of Fig.~\ref{typ_disk4} the evolution of the angle $\theta_{i_1^2}=5\lambda_2-2\lambda_1-\varpi_1-2\Omega_1$ with the evolution of the mutual inclination (in logarithmic scale). We see that the inclination increases in correlation with the libration of the inclination-type resonant angle $\theta_{i_1^2}$.

\begin{figure}[]
	\centering
	\includegraphics[width=0.49\columnwidth]{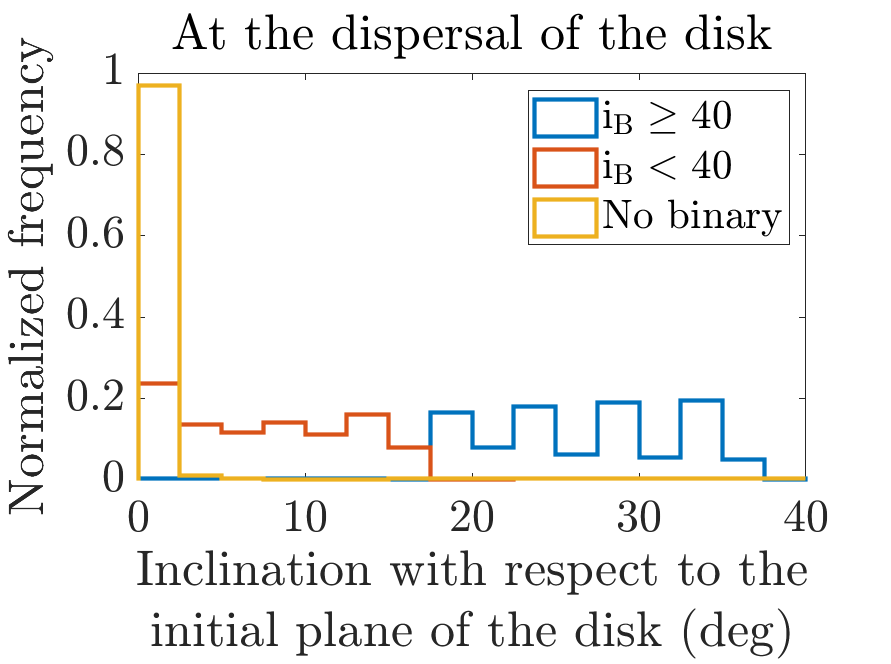}
	\includegraphics[width=0.49\columnwidth]{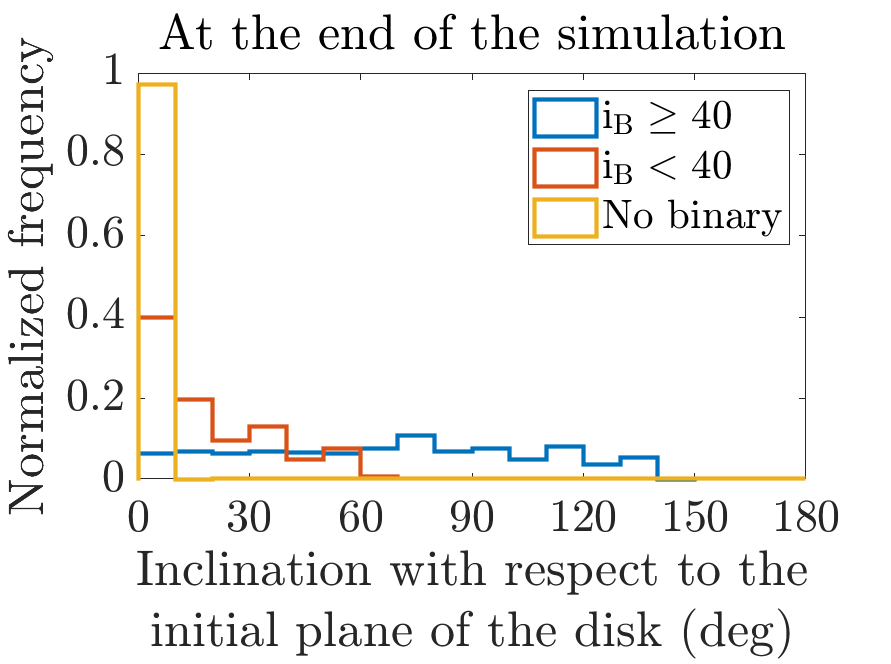}
	\caption{Inclination distribution at the dissipation of the disk (left panel) and at the end of the simulation (right panel). The inclinations are given in the initial disk plane reference frame.}
	\label{i2pl}
\end{figure}

The planets mostly remain in the midplane of the disk during the disk phase thanks to the gravitational and damping forces of the disk. We note that the planets tend to be slightly more misaligned with respect to the disk in systems with a binary companion but this misalignment is not significant (less than $4^\circ$).

In Fig.~\ref{i2pl}, we display the planetary inclinations with respect to the initial plane of the disk. We observe in the left panel showing the inclinations at the dispersal of the disk, that for binary star systems the planetary inclinations are proportional to the inclination of the binary companion. This is coherent with the precession of the disk due to the binary companion, which induces a periodic variation with an amplitude equal to $2i_B$. Since the planets tend to stay coupled with the disk, they follow its nodal precession, which explains the inclination distribution in the left panel of Fig.~\ref{i2pl}.
On the right panel, we observe that the inclination is amplified during the dynamical phase. This is a consequence of the nodal precession induced by the binary companion acting this time on the coupled planets. Moreover, on the right panel, we observe that the normalized frequencies decrease with the inclination. This effect comes from the amplitude of the nodal precession. Indeed, we have 400 systems for each inclination of the binary, whose planetary inclinations are uniformly spread between 0 and $2i_B$, leading to an accumulation around the smallest values.

Assuming that the disk angular momentum vector is originally aligned with the stellar spin and that the stellar spin does not change direction over time \citep[although see][for conditions where this assumption might break down]{Storch_2014}, and observing that the planets mostly remain in the disk's midplane throughout the migration, we can use the disk's inclination as a proxy for measuring spin-orbit angles.
The nodal precession induced by the binary companion thus creates a misalignment between the spin axis of the primary star and the angular momentum of the planets.

\begin{figure}[]
	\centering
	\includegraphics[width=0.49\columnwidth]{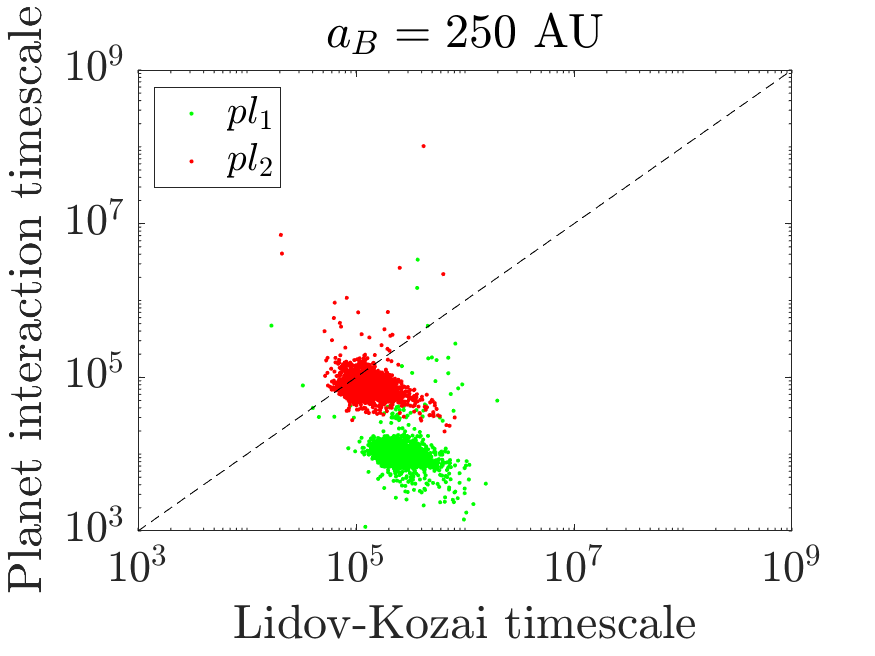}
	\includegraphics[width=0.49\columnwidth]{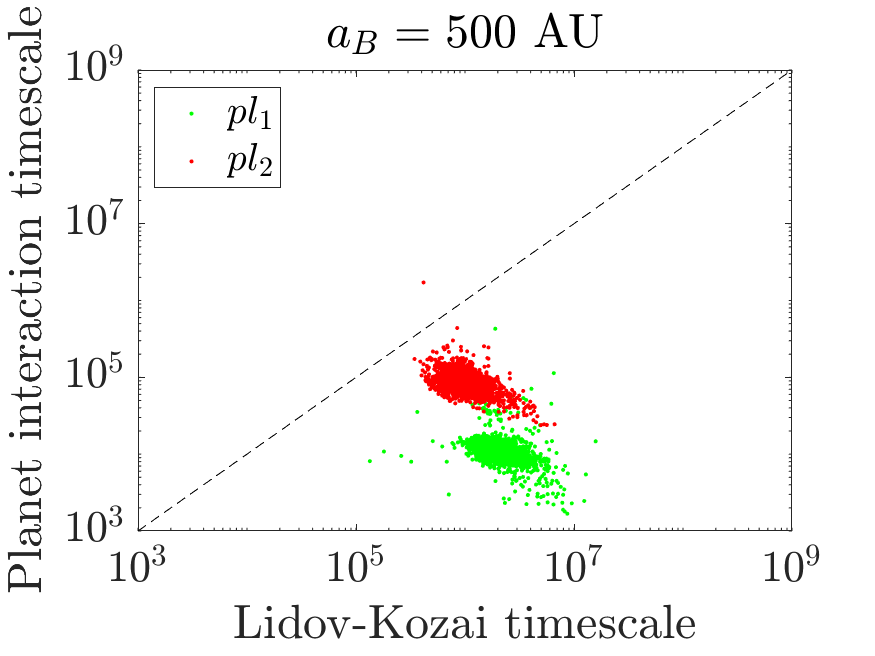}
	\includegraphics[width=0.49\columnwidth]{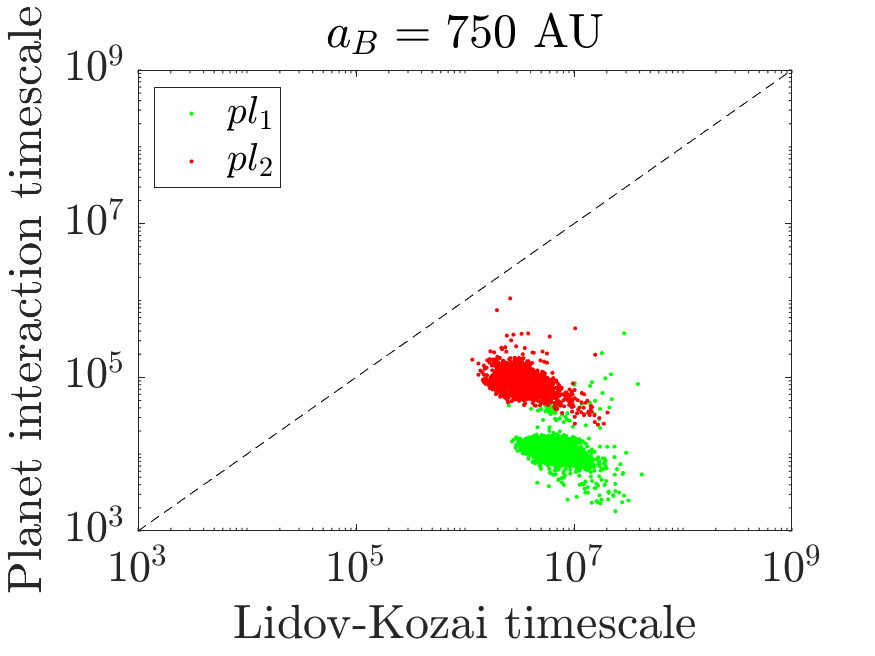}
	\includegraphics[width=0.49\columnwidth]{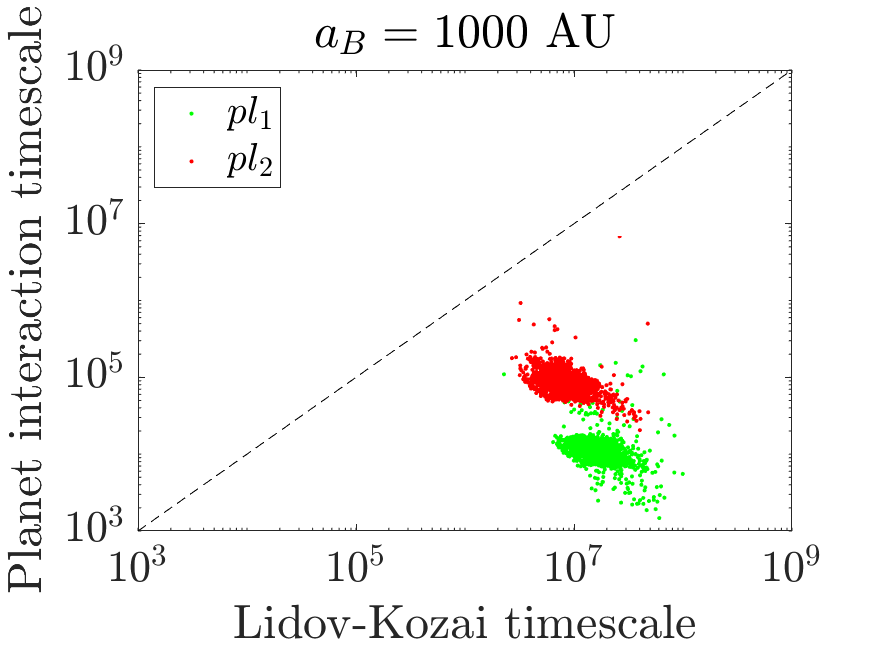}
	\caption{Same as Fig.~\ref{freq_comp} for the different semi-major axes of the binary companion.}
	\label{freq_comp_astar2}
\end{figure}

\subsubsection{ZLK resonance}

To further analyze the influence of the binary companion, in particular the ZLK mechanism, we now focus on the pericenter argument of the planets. The bottom panels of Fig.~\ref{P_ratio} display the pericenter argument distributions at the dissipation of the disk (left panel) and at the end of the simulation (right panel). We do not notice any specific pattern associated with the ZLK resonance (i.e., accumulations around 90 or 270$^\circ$). Regarding the evolution after the disk phase (right panel), the gravitational interaction between the two planets tends to overcome the effect of the binary companion and the establishment of a ZLK resonant evolution in highly inclined binary systems. Note that we also numerically investigated the libration of the pericenter arguments and observed that only the systems suffering from an ejection (i.e., systems with one remaining planet) possibly hold a planet locked in the ZLK resonance.

Overall, our simulations indicate that for planetary systems formed in very wide binaries, resonant captures still occur frequently and are robust against perturbations from the binary. In particular, on the long-term evolution, resonant interactions seem to quench ZLK oscillations.

\begin{figure}[]
	\centering
	\hspace{0cm}
	\vspace{0.25cm}
	\includegraphics[width=0.49\columnwidth]{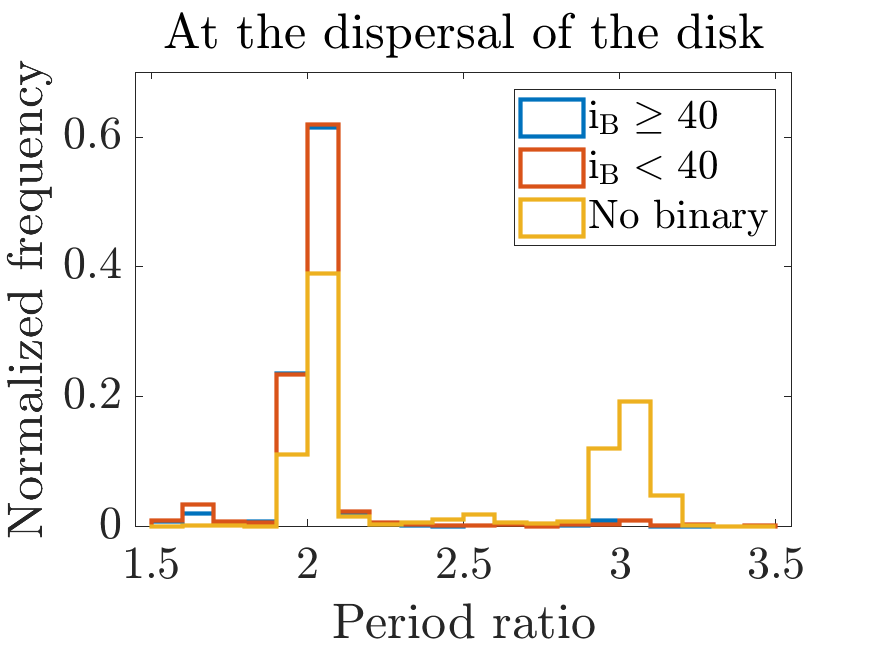}
	\includegraphics[width=0.49\columnwidth]{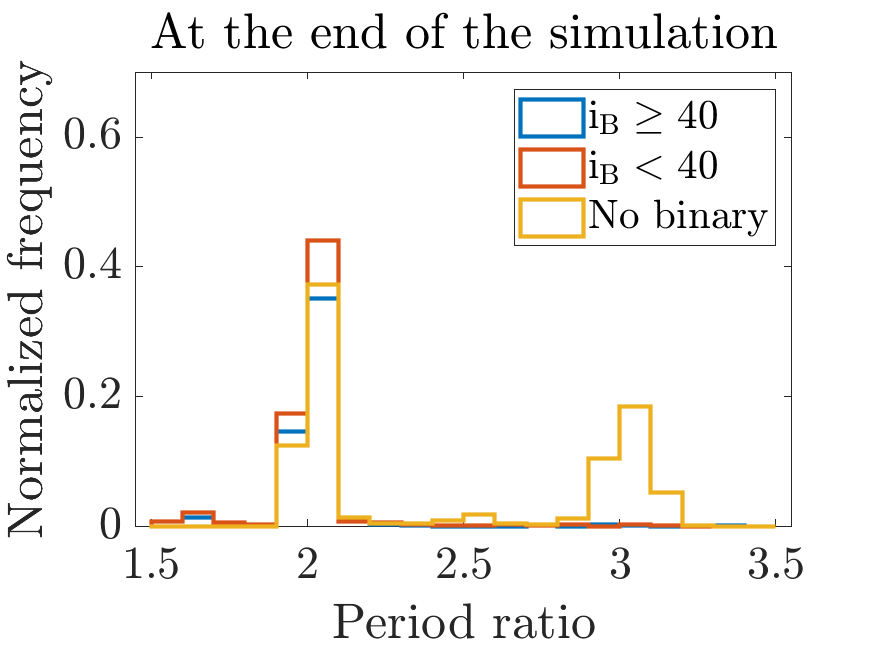}
	\vspace{0.25cm}
	\includegraphics[width=0.49\columnwidth]{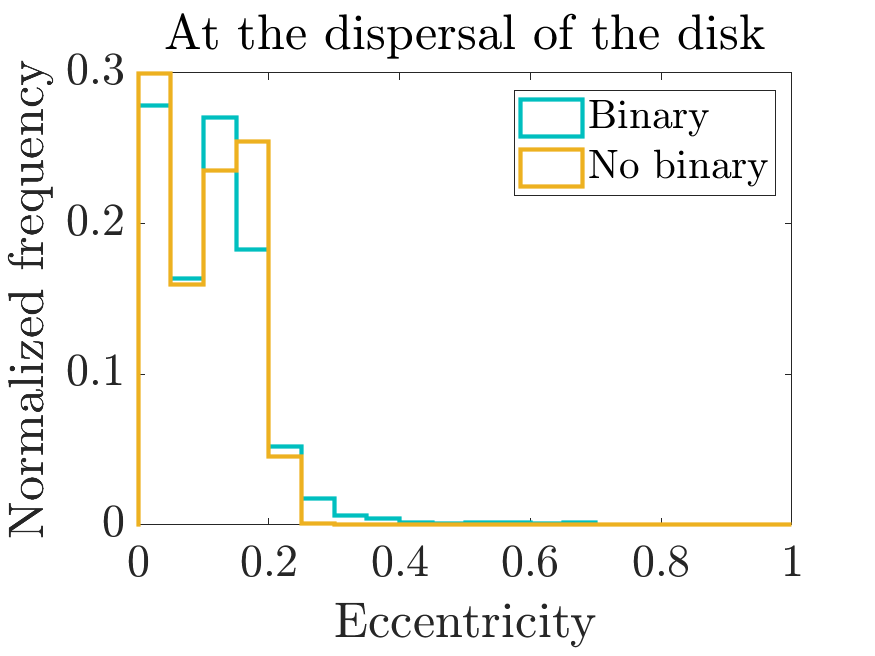}
	\includegraphics[width=0.49\columnwidth]{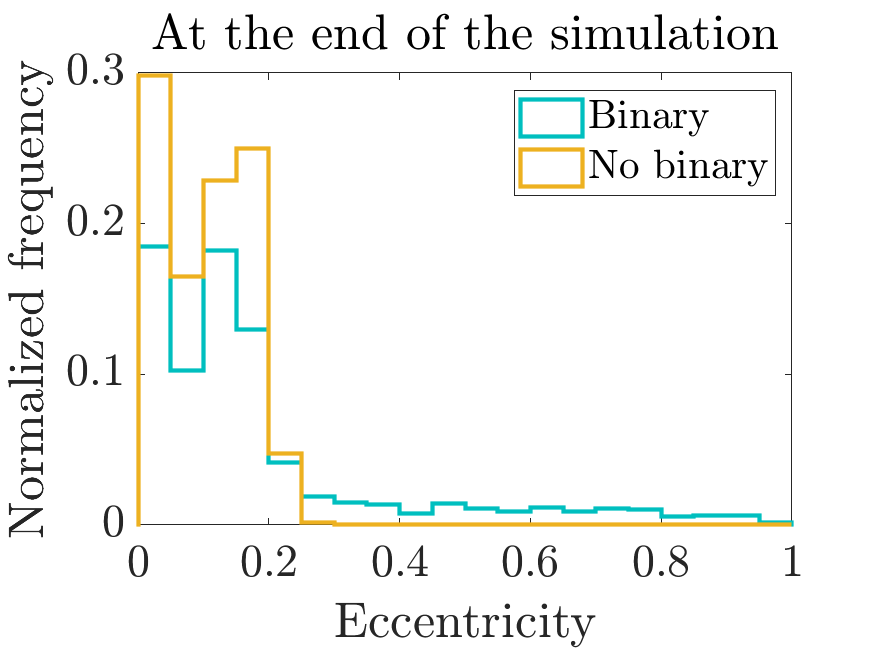}
	\vspace{0.25cm}
	\includegraphics[width=0.49\columnwidth]{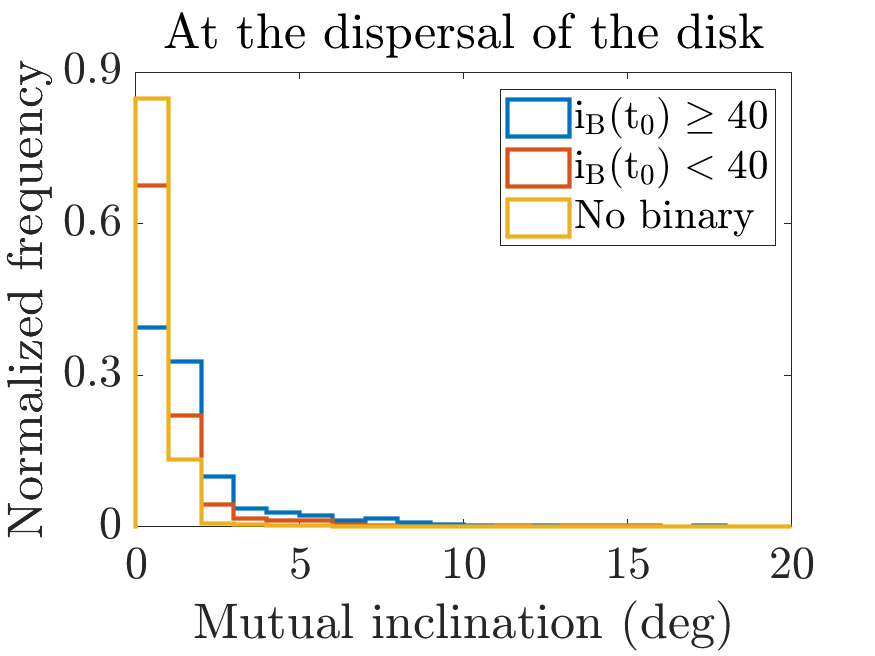}
	\includegraphics[width=0.49\columnwidth]{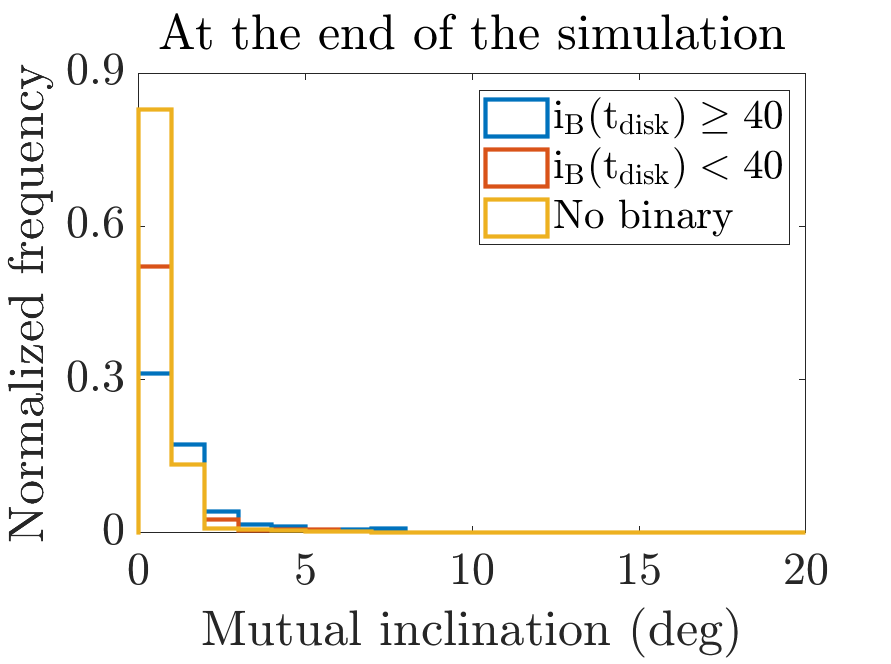}
	\includegraphics[width=0.49\columnwidth]{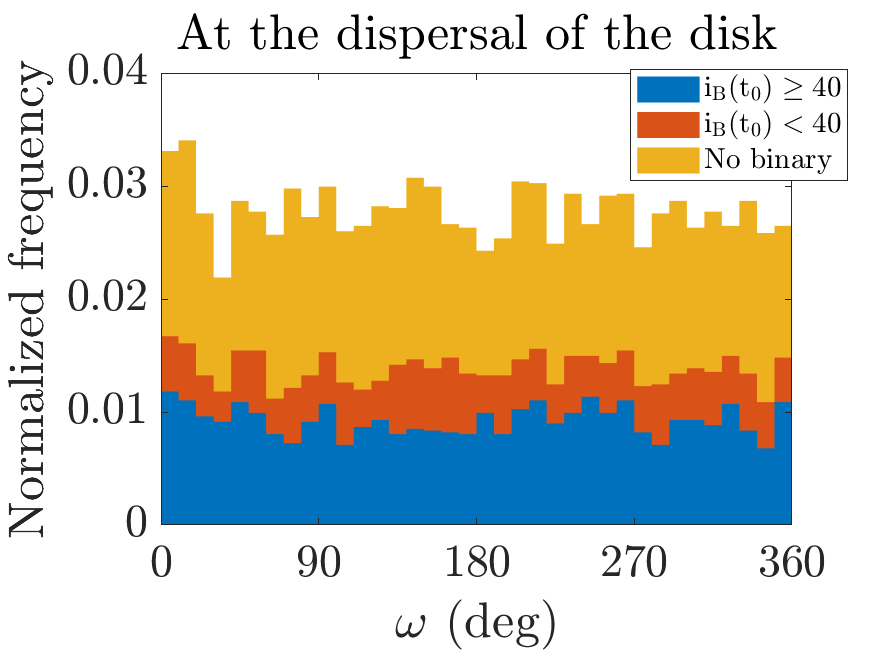}
	\includegraphics[width=0.49\columnwidth]{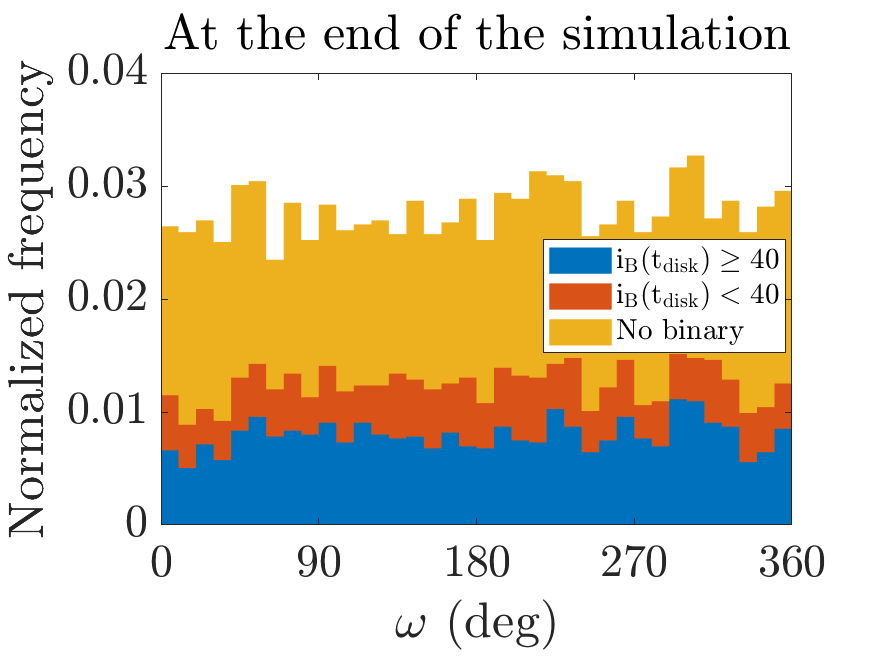}
	\caption{Same as Fig.~\ref{P_ratio}, for $a_B=250$ au.}
	\label{e_250}
\end{figure}

\section{Closer binary with $a_B=250$ au}
\label{sec:close_bin}
As seen in the previous section, in case of a wide binary companion at $1000$~au, the influence of the binary companion on planet pairs embedded in the gas disk is very limited and the resonant evolution of the pairs is preserved well after the dispersal of the disk. In this section, we investigate the case of closer binary stars.


\subsection{Second set of simulations}
Based on the observations related to Fig.~\ref{freq_comp}, we aimed to conduct a parametric study to identify parameter values leading to an increase of the ratio between the precession timescale $\tau_{\rm pl}$ and the ZLK resonance timescale $\tau_{\rm LK}$. Following Eq.~\eqref{tau_LK}, to reduce the ZLK timescale, we decreased the semi-major axis of the binary companion by adopting the values 250~au, 500~au, and 750~au. Note that when the binary companion is closer to the primary star, our disk model suffers from some limitations that will be discussed in detail in Section~\ref{Limitations}. We also increased the initial semi-major axes of the planets by multiplying their range of values by 1.5 in order to increase their orbital periods. A last change consisted in decreasing the lower bound on the planetary masses to $0.65~M_{\rm Jup}$. All the other parameters were kept unchanged, as given in Table~\ref{Body_param2}. For each of the four $a_B$ values and for the 32 choices of the binary eccentricity and inclination, we randomly drew 50 different initial conditions for the planetary parameters. In total, we ran 8000 simulations (single star simulations included).

We now study the system configurations resulting from this second set of simulations at the dispersal of the disk. In Fig.~\ref{freq_comp_astar2}, we reproduce the same plot as in Fig.~\ref{freq_comp} for the different semi-major axes of the binary companion considered here. As expected from Eqs.~\eqref{tau_LK}, \eqref{eig_+}, and \eqref{eig_-}, we see that the smaller the semi-major axis of the binary, the closer the systems to the dashed line of equal timescales. As a result, the impact of the binary companion will be stronger for simulations with $a_B=250$~au and we therefore chose to focus on this binary separation in the remainder of this section.

\subsection{Results}

\begin{figure}[]
	\centering
	\hspace{0cm}
	\vspace{0.5cm}
	\includegraphics[width=0.49\columnwidth]{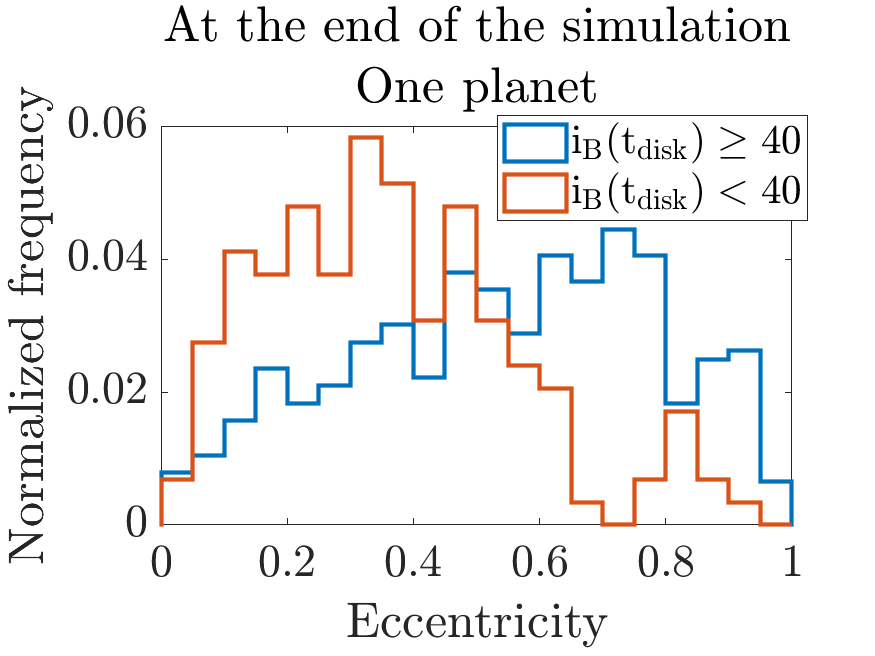}
	\includegraphics[width=0.49\columnwidth]{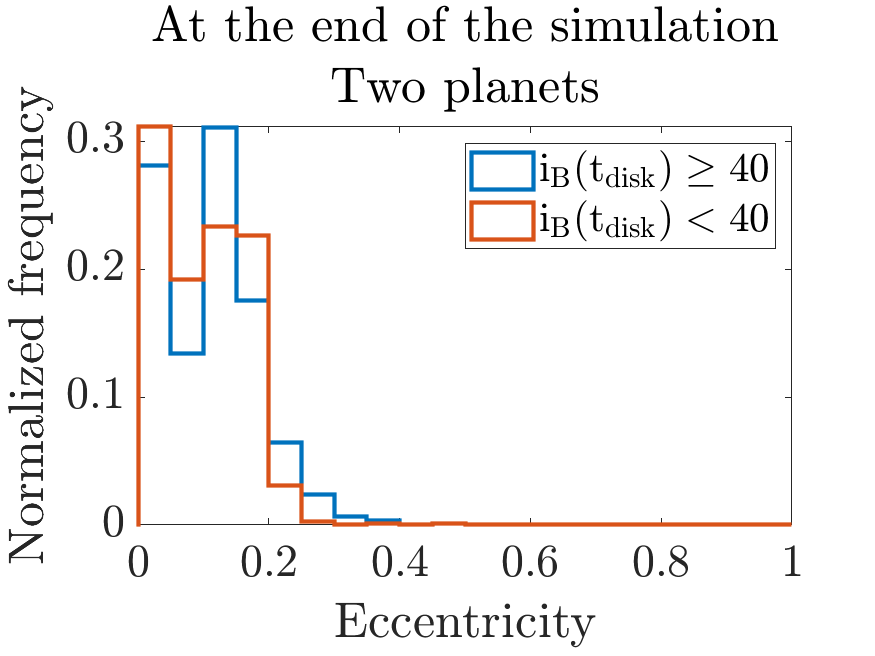}
	\includegraphics[width=0.49\columnwidth]{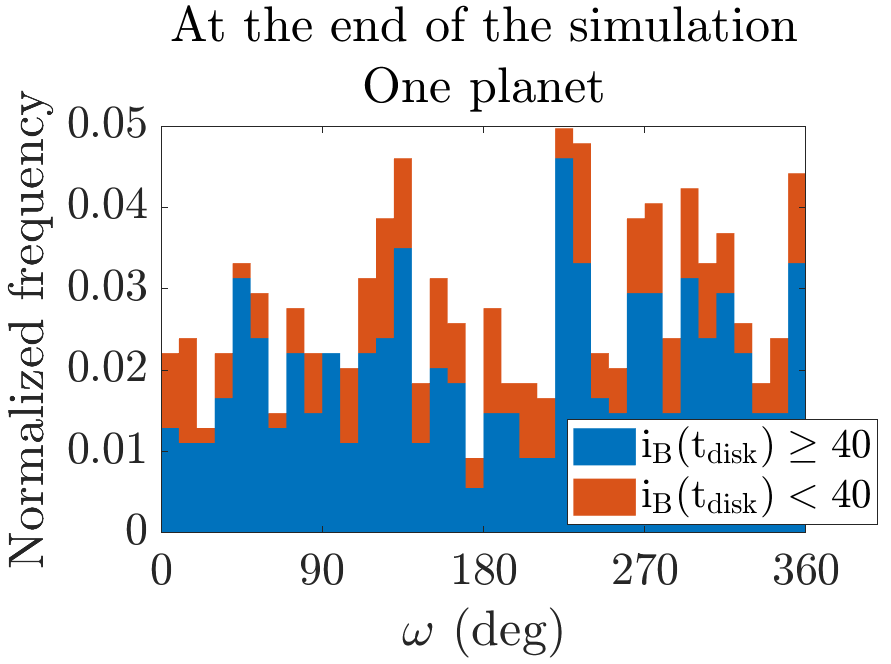}
	\includegraphics[width=0.49\columnwidth]{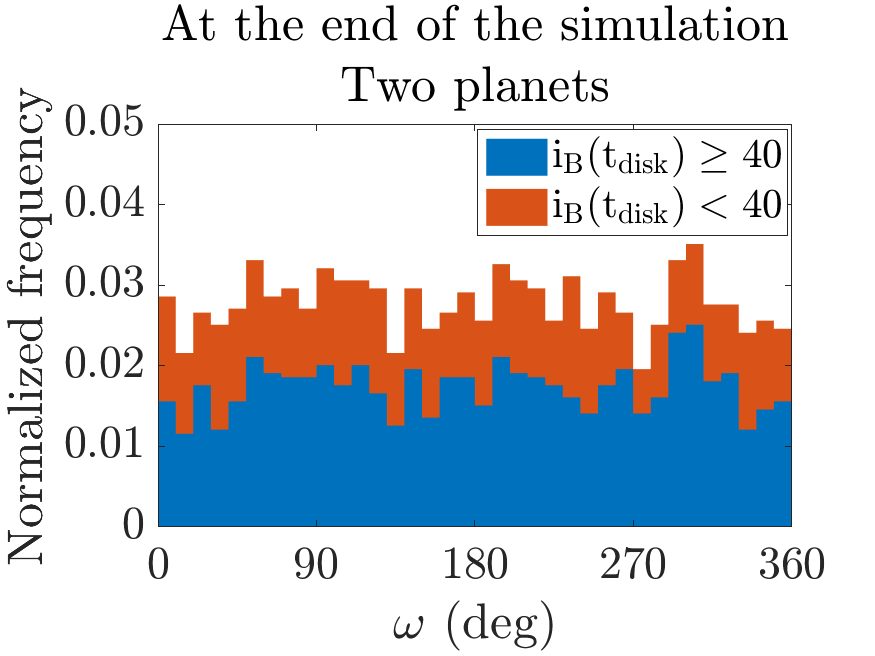}
	\caption{Eccentricity and pericenter argument distributions at the end of the simulations for single planet systems (left panels) and two-planet systems (right panels) when $a_B=250$ au}
	\label{e_250_pl}
\end{figure}

The results are displayed in Fig.~\ref{e_250}. We observe that systems with a binary companion (cyan color) are more eccentric than the ones without companion (yellow color), especially at the end of the simulations where highly eccentric planetary orbits are observed. They mainly result from scattering events and ejections during the long-term evolution. As shown in Fig.~\ref{e_250_pl}, for the simulations with a binary companion ending with two planets, almost all the eccentricities are below 0.2 (right panel), while most of the one-planet systems present very eccentric configurations (left panel). Note that in Fig.~\ref{e_250_pl} we also see that the inclination of the binary companion influences the planetary eccentricities, since we observe that the majority of the highly eccentric systems have a binary companion with an inclination higher than $40^\circ$ (i.e., close to the critical value of $39.23^\circ$ for the ZLK mechanism.).

Again, the mutual inclinations between the planets are small, mainly below $4^\circ$ (panels of the third row of Fig.~\ref{e_250}). Moreover, the distributions of the pericenter arguments shown in the bottom panels of Fig.~\ref{e_250} are nearly uniform, but we note this time a possible evidence of the ZLK resonance. Indeed, if we present separately the single planet systems and the two-planet systems formed at the end of the simulations, we see in Fig.~\ref{e_250_pl} that the distribution of the systems ending with one planet only presents two peaks centered at $90^\circ$ and $270^\circ$  associated with the ZLK resonance (bottom left panel). A study of the libration of the pericenter arguments for each system individually leads us to the observation that no system with two planets end up with one of the planets in the ZLK resonance with the binary companion. For the one-planet systems, the percentage of systems locked in the ZLK resonance with an inclined binary companion is $\sim30\%$, similarly to the observation made in \cite{Roisin_2021b}. However, even if the system is not evolving in the ZLK resonance, its dynamics is still influenced by the ZLK islands which can explain the high planetary eccentricities observed here. For a dynamical study of the one-planet case, we refer to \cite{Roisin_2021a,Roisin_2021b}.

\begin{figure}[]
	\centering
	\includegraphics[width=\columnwidth]{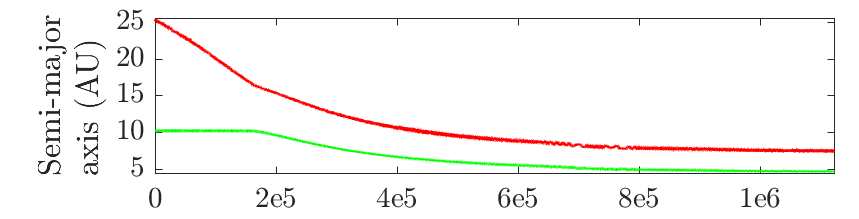}
	\includegraphics[width=\columnwidth]{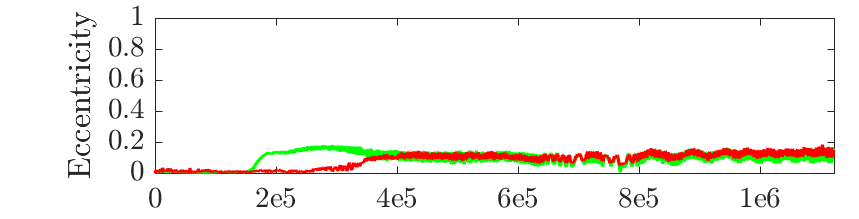}
	\includegraphics[width=\columnwidth]{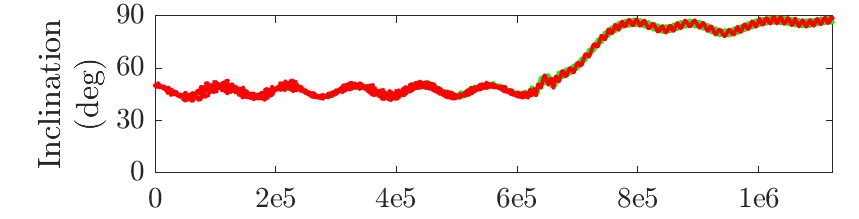}
	\includegraphics[width=\columnwidth]{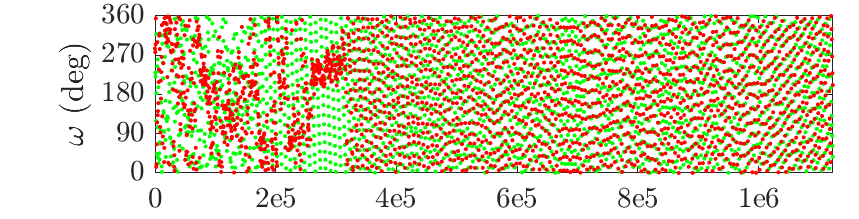}
	\includegraphics[width=\columnwidth]{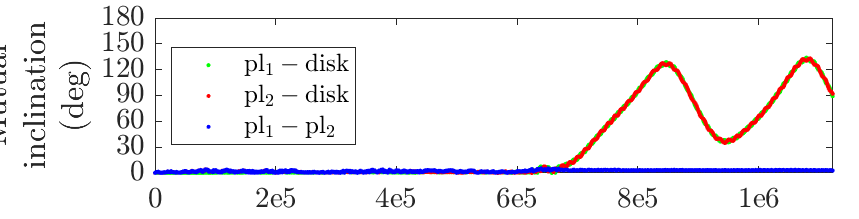}
	\includegraphics[width=\columnwidth]{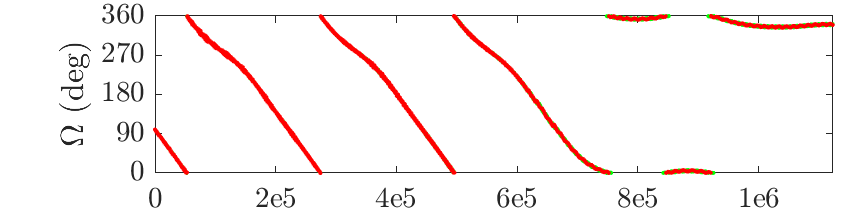}
	\includegraphics[width=\columnwidth]{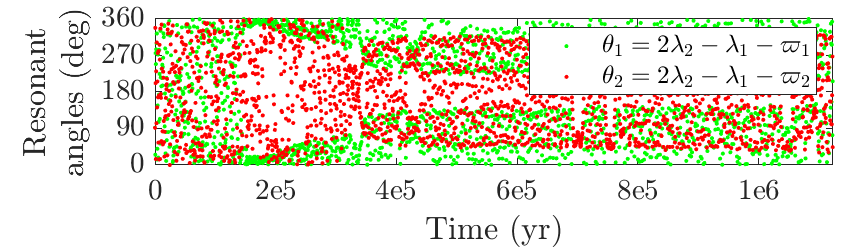}
	\caption{ Typical evolution of a system of two giant planets with a binary at $a_B=250$ au (in the Laplace plane reference frame), during the disk phase. The initial parameters (with respect to the disk plane) are $e_B=0.5$ and $i_B=50^\circ$ for the binary companion and $a_{1}=10.22$ au, $m_{1}=1.39~M_{\rm Jup}$, $a_{2}=25.33$ au, and $m_{2}=1.07~M_{\rm Jup}$ for the planets.}
	\label{typ_250}
\end{figure}

\begin{figure}[]
	\centering
	\includegraphics[width=0.6\columnwidth]{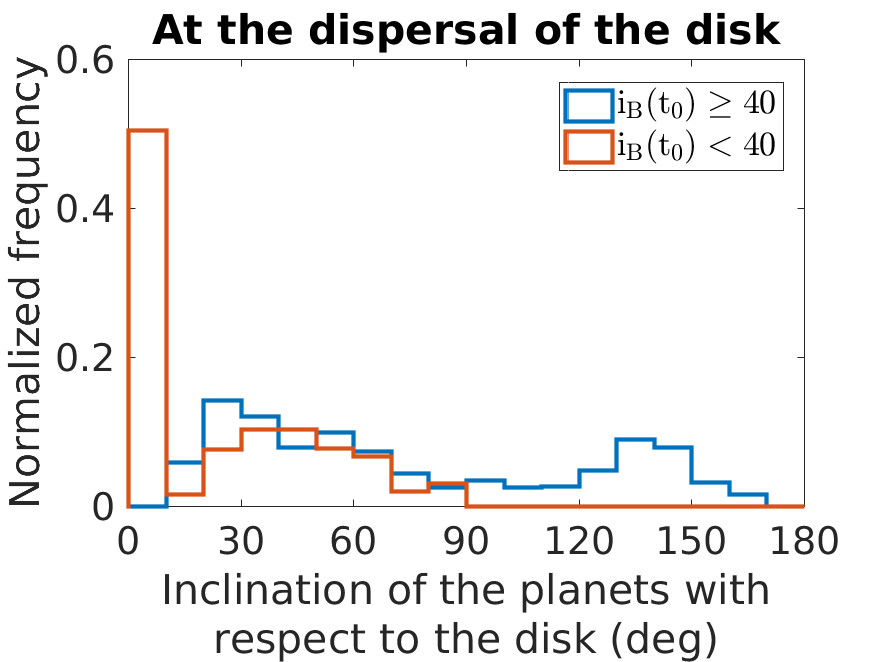}
	\caption{Histogram of the inclination of the planets with respect to the disk, at the dissipation of the disk for simulations with a binary separation $a_B=250$~au. Blue (resp. orange) bins are for binary orbital inclinations above (resp. below) $40^\circ$. }
	\label{i_250}
\end{figure}

Contrary to the case where $a_B=1000$~au, the inclination of the planets with respect to the disk was modified during the disk phase for $a_B=250$~au, as a consequence of the much shorter timescale of the binary perturbations. This can be observed in the typical evolution displayed in Fig.~\ref{typ_250}. In Fig.~\ref{i_250} we show the inclination of the planets with respect to the disk at the time of the dispersal of the disk. For binaries with inclinations less than $40^\circ$ (red color), about half of the planets remain in the disk's midplane, while the other half show inclinations ranging up to $90^\circ$, with a peak around $40^\circ$. This distribution could be in part caused by the nodal precession which induces inclinations ranging from 0 to $2i_B$. Indeed, some of these planets may have been separated from the disk's midplane due to the differential precession induced by the binary, which occurs on a timescale short enough that the damping forces and disk-induced precession cannot counteract it. Note that some planets may also have been expelled from the disk's midplane due to inclination-type resonances, as illustrated in Fig.~\ref{typ_disk4}. For binaries with inclinations larger than $40^\circ$ (blue color), the distribution of the mutual disk-planet inclinations broadens and an excess is observed at inclinations near $30^\circ$ and $140^\circ$. After being ejected from the disk's midplane, these planets may be undergoing ZLK cycles driven by the disk, as shown by \cite{Terquem_2010}, for which the critical angle can be lower than $40^\circ$ \citep{Teyssandier_2013}.

\begin{table}[]
	\caption{Percentages of the different system configurations at the end of the simulation for the second set of simulations}
	\hspace{-0cm}
	\begin{tabular}{lrrrrr}
		\hline
		& $250$ au & $500$ au & $750$ au & $1000$ au  & No  binary \\
		\hline
		2 planets & & & & & \\
		\quad MMR & 57 & 89.44 & 90.31 & 90.5 & 91.12\\
		\quad no MMR& 5.25 & 7.44 & 8.31 & 7.87 & 7.94\\
		1 planet & 33.94 & 2.56 & 0.82 & 0.94 & 0.63 \\
		no planet & 3.81 & 0.56  & 0.56 & 0.69 & 0.31\\
		\hline
	\end{tabular} 
	\label{astar2}
\end{table}

\begin{figure}[]
	\centering
	\includegraphics[width=0.9\columnwidth]{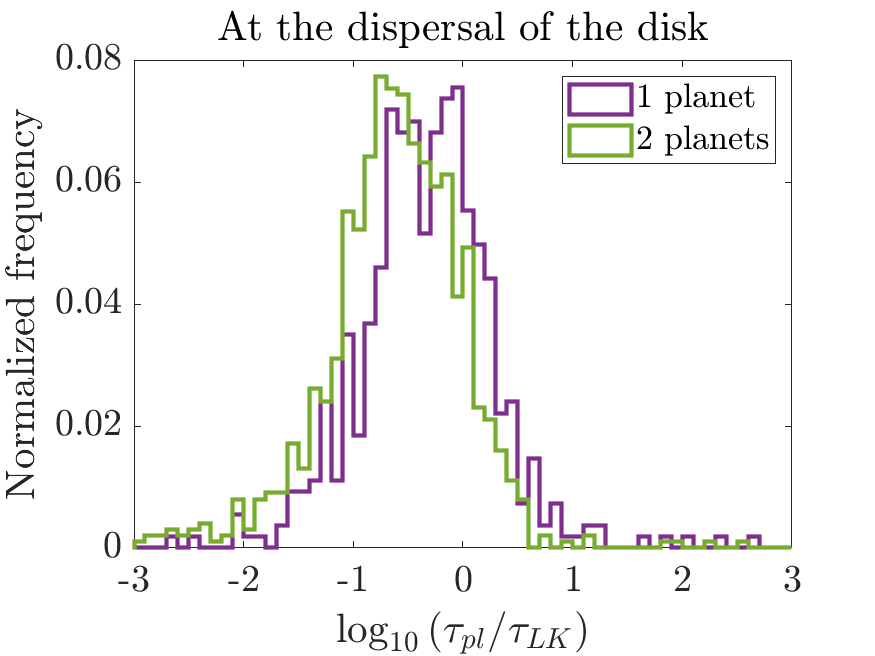}
	\caption{Normalized frequency of the logarithm of the ratio between the timescale of the gravitational interaction between the planets and the ZLK timescale, for the simulations with $a_B=250$ au. The color code refers to the number of planets in the system at the end of the simulation.}
	\label{Freq_ratio}
\end{figure}

Regarding the MMRs, a variation in the initial semi-major axes of the planets leads to a modification of their migration rates and thus possibly to different MMR captures. In the new set of simulations with smaller $a_B$ values, almost all the resonant systems are trapped in the 2:1 MMR. During the disk phase, the evolutions of the systems are very similar to the ones with a wide binary companion described in Section~\ref{sec:wide_bin}. At the end of the simulation, a lot of ejections are reported for the simulations with $a_B=250$ au, as can be observed in Table \ref{astar2} which shows the system configurations at the end of the simulation for the four considered binary separations for comparison. In Fig.~\ref{Freq_ratio}, we observe that, when $a_B=250$ au, the systems ending with a single planet at the end of the simulation (purple curve) have a mean ratio between $\tau_{\rm pl}$ and $\tau_{\rm LK}$ higher than the one for systems with two planets (green curve). This confirms that systems for which both timescales are roughly similar experience more planetary ejections. For all the other values of the binary semi-major axis, Table \ref{astar2} shows that the influence of the binary companion on the system architecture is still quite limited.

In conclusion, we observed that a binary companion close enough to strongly influence the planets ($a_B=250$ au) will tend to destabilize the systems and lead to the ejection of a planet in about one third of the cases. The ejection will possibly induce high eccentricity and ZLK resonance capture for the remaining planet.

\begin{figure}[]
	\centering
	\includegraphics[width=0.9\columnwidth]{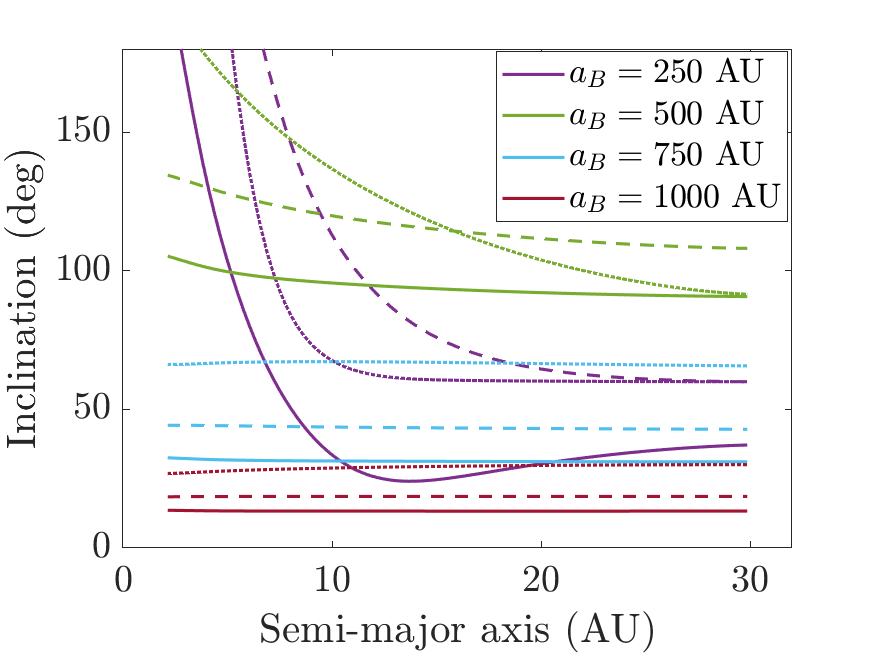}
	\caption{Time evolution of the inclination of the disk as a function of the semi-major axis, for different semi-major axes of the binary (in the initial disk plane reference frame). The line type refers to different times: the solid line stands for t=0.5 Myr, the dashed line for t=0.75 Myr and the dotted line for t=1.123 Myr. The inclination of the binary $i_B$ is fixed to $60^\circ$.}
	\label{a_B_influence}
\end{figure}

\subsection{Caveats of the simulations with $a_B=250$~au}
\label{Limitations}
In this section we discuss the possible limitations of our disk model for the parameters of the second set of simulations. Indeed, we have assumed here that the disk is uniform due to its self-gravity. This hypothesis is valid for wide binary stars \citep[e.g.,][]{Roisin_2021b}. Closer binary companions as the ones considered in the second part of our work could have the effect to overcome the disk self-gravity and lead to deformations of the disk. By splitting the disk in several massive rings adjacent to one another and making them evolve using the Laplace-Lagrange theory to estimate the nodal precession of the disk, we studied in Fig.~\ref{a_B_influence} the uniformity of the disk for the different semi-major axes of the binary companion considered here. More details about this technique can be found in \cite{Murray_1999,Levison_2007,Batygin_2011,Roisin_2021b}. Note that we fix the inclination of the binary companion to $i_B=60^\circ$ arbitrarily, but the results are similar regardless of the chosen inclination. We see that the disk remains uniform for a binary companion at 750~au and~1000 au. On the contrary, the disk uniformity is not maintained for 250~au and 500~au. 
This does not mean that the disk is not uniform in this case, but the self-gravity is not sufficient alone to ensure it. Hydrodynamical effects such as radial pressure force and viscous diffusion \citep[e.g.,][]{Papaloizou_1995,Larwood_1996} could play a role and keep the disk uniform. For instance, it has been shown that if the wave crossing time in the disk is shorter than the precession time due to the binary companion, the rigidity of the  disk could be maintained \citep{Zanazzi_2018a}. Investigations of such effects is left for future work.

Moreover, the disk could experience ZLK cycles due to the presence of the binary companion \citep[see, e.g.,][]{Martin_2014, Fu_2015, Zanazzi_2017}. \cite{Batygin_2011} showed that the cycles could be suppressed by the self-gravity of the disk if the disk is massive enough. This assumption will eventually break down in our simulations since we implemented an exponential decrease of the disk mass. In addition, ZLK cycles in the disk could lead to eccentric orbits and the hypothesis of the Laplace-Lagrange theory requiring annuli without intersections could not be verified in this case.

Nevertheless, most of the results of the second set of simulations were very similar for all the semi-major axes considered for the binary companion. The differences observed for $a_B=250$ au mainly arise during the long-term evolution after the dispersal of the disk. Since this phase does not suffer from the limitations in the disk model presented in this section, we therefore conclude that despite those limitations, our results are quite robust.

\begin{figure*}[]
	\centering
	\includegraphics[width=1.8\columnwidth]{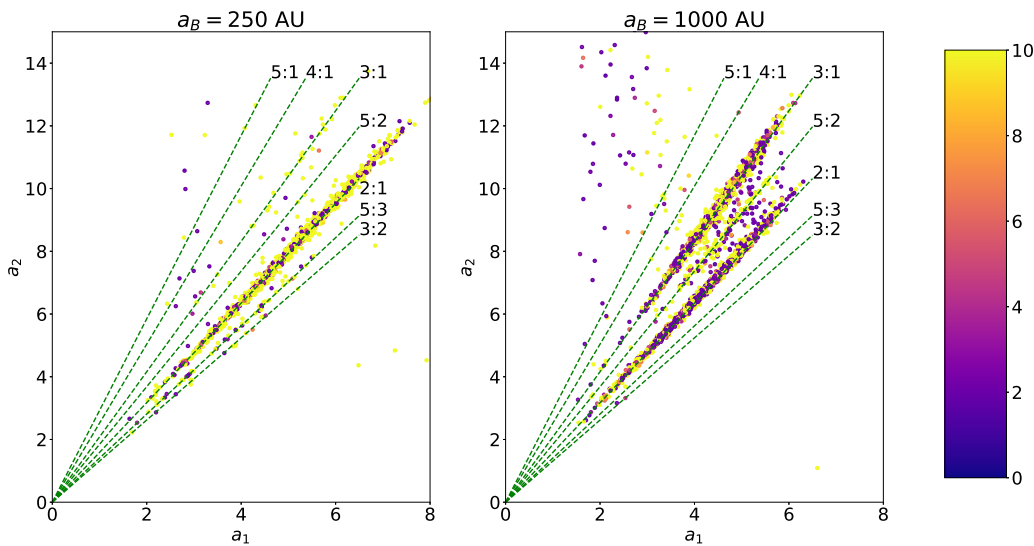}
	\caption{Semi-major axis of the outer planet with respect to the one of the inner planet for the two-planet systems formed in our simulations with the binary at $a_B=250$~au (left panel) and at $a_B=1000$~au (right panel). The color code refers to the MEGNO value.}
	\label{MEGNO_MMR_graph}
\end{figure*}

\begin{table*}
	\centering
	\caption{Percentages of regular and irregular evolutions as given by the MEGNO indicator. The percentages are shown for different inclinations of the binary and MMRs between the planets. See the text for more details.}
	\begin{tabular}{cccccccccc}
		\hline
		MEGNO & \multicolumn{2}{c}{$a_B=250$ au} & \multicolumn{2}{c}{$a_B=1000$ au} &  {$a_B=250$ au} & \multicolumn{3}{c}{$a_B=1000$ au}   \\
		&  $i_B<40^\circ$ & $i_B\geq 40^\circ$ &  $i_B<40^\circ$ & $i_B\geq40^\circ$ & 2:1 & 2:1 & 5:2 & 3:1  \\
		\hline
		$[0:2.5]$ & 26.3 & 16.7 & 52.2 & 44.7  & 22 & 52.3 & 24.2 & 39.2\\
		$[2.5,6]$& 7.8 & 6.7 & 10.2 & 11 & 7.5 & 9.8 & 7 & 13.7 \\
		$>6$& 65.9 & 76.6 &  37.6 & 44.3 & 70.5 & 37.9 & 68.8 & 47.1\\
		\hline
	\end{tabular} 
	\label{MEGNO}
\end{table*}

\section{Stability}
\label{sec:stability}

To assess the stability of the two-planet systems formed in our simulations, we used the Mean Exponential Growth factor of Nearby Orbits (MEGNO) chaos indicator which is based on the evolution of one deviation vector \citep{Cincotta_2003}. This indicator differentiates efficiently the orbits since it converges to 0 for stable periodic orbits, to 2 for quasi-periodic orbits and orbits close to stable periodic orbits, and diverges with time for irregular orbits. In practice, the simulations were made with the symplectic Wisdom-Holman integrator WHFast implemented in REBOUND \citep{reboundwhfast} and we started to follow the evolution of the MEGNO indicator just after the disk phase.

A visual representation of the results is given in Fig.~\ref{MEGNO_MMR_graph} which shows for each system the semi-major axes of the two planets. The two-planet systems with a binary companion at $250$~au are shown in the left panel and the ones when $a_B=1000$~au are shown in the right panel. The green dashed lines indicate the different MMRs. The high density of points close to the lines confirms that the systems gather around the 2:1 MMR for closer binaries, while captures in higher order resonances like the 3:1 and 5:2 MMRs are possible for wider binaries. Moreover, as previously noted, fewer two-planet systems are formed when the binary companion is at $250$~au due to the largest proportion of planetary ejections.

The color code of Fig.~\ref{MEGNO_MMR_graph} refers to the MEGNO value. Note that the color bar upper value is fixed to 10 since we stopped the simulation when the MEGNO reached this value. We observe that in the wide binary case, many systems close to the 2:1 and 3:1 MMRs have a regular motion while the evolution of most of the systems near the 5:2 MMR is irregular. The exact percentages of systems with regular (defined here as a MEGNO value smaller than 2.5) and irregular (MEGNO value higher than 2.5) is given in Table~\ref{MEGNO}. On the contrary, for close binaries, only $22\%$ of the systems near the 2:1 MMR present a regular evolution, showing the significant impact of a close binary companion on the long-term evolution of planetary systems. As expected, we observe in Table~\ref{MEGNO} that the percentage of regular systems is slightly lower for highly inclined binary stars (above $40^\circ$).

%


\section{Comparison with observations}
\label{sec:obs}

\begin{figure}[]
	\centering
	\includegraphics[width=\columnwidth]{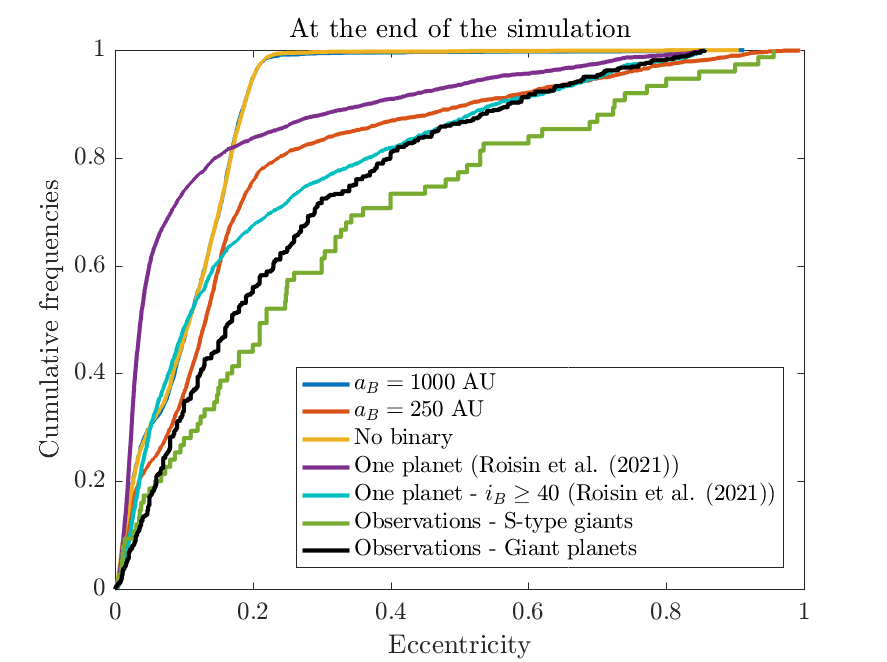}
	\caption{Cumulative eccentricity of the planets found in our simulations and in the observations \citep{Rein_2012}. See text for more details.}
	\label{e_freq_obs}
\end{figure}

In this section we compare our results with the 187 S-type planets in binary star systems presented in the Open Exoplanet Catalogue of \cite{Rein_2012}. In particular, the comparison of the cumulative eccentricities is made in Fig.~\ref{e_freq_obs}. The eccentricity distribution of \cite{Rein_2012} for S-type planets is indicated by the green curve, while the one of all the detected planets is indicated by the black curve. Note that we only considered Jupiter-like planets with a mass higher than $0.5 M_{\rm Jup}$. As noticed in \cite{Kaib_2013}, planets in S-type systems have larger eccentricities. We observe that for the systems with a binary companion at 1000 au (blue curve) and without binary companion (yellow curve) found in our simulations, nearly all the eccentricities are below $\sim0.2$, since the two-planet systems are weakly perturbed during their long-term evolution. On the contrary, high eccentricities are found in simulations with a close binary companion at 250 au (orange curve) and can notably be explained by the influence of the ZLK resonance on the planetary evolution after the ejection of one of the planets \citep[see][for more details]{Roisin_2021a}. We also added in Fig.~\ref{e_freq_obs} the systems found in \cite{Roisin_2021b} for simulations of single planets migrating in the disk with a binary companion at 1000 au (purple curve). While binary companions with $i_B<40^\circ$ do not excite the initial quasi-circular orbits of the planets, high eccentricities can be reached in the presence of a highly inclined binary companion (cyan curve). As a result, we showed that the high eccentricities observed for S-type planets in Fig.~\ref{e_freq_obs} can be caused by highly inclined wide binary companions in case of single planet systems and close binary companions in case of multi-planet systems.

Moreover, several observations of misalignment between the stellar spin axis and the orbital angular momentum of hot Jupiters have been reported \citep{Wright_2011,Albrecht_2012}. A common scenario to explain this misalignment is the action of tidal friction during the ZLK cycles due to an highly inclined binary companion \citep[e.g.,][]{Naoz_2014}. In this work we observed misalignment induced by the nodal precession of the planets due to their interaction with the binary companion, regardless of the inclination of the binary companion. This effect was extensively studied during the disk phase \citep[see, e.g.,][]{Batygin_2013,Lai_2014,Spalding_2014,Zanazzi_2018b} but it continues even after the dissipation of the disk \citep[][]{Boue_2014}. The nodal precession effect could lead to possibly strong misalignment even for planets further away of the star than the hot Jupiters and could help explaining the observed misalignment, alongside other mechanisms. Note that in this work we considered the spin axis of the star as fixed and parallel to the disk angular momentum at the beginning of the simulation. Of course, it is a strong assumption since the disk would most likely precess before the formation of the giant planets and would be misaligned. Moreover, the spin axis of the host star could also be chaotic and not constant, as shown by \cite{Storch_2014} for a primary star under the influence of a hot Jupiter for example.

\section{Conclusions}
\label{sec:conclusion}
In this paper, we studied the impact of a binary companion on the migration of two giant planets in a protoplanetary disk, followed by the long-term evolution of the systems after the disk has vanished. The eccentricity and inclination damping induced by the disk, its gravitational potential, and the nodal precession induced by the binary companion on the disk were all considered here. A timescale analysis hinted at the existence of two main patterns in the disk migration phase. First, for very wide binaries ($a_B=1000$~au), the nodal precession induced by the binary occurs on a timescale longer than the disk's lifetime. In this case, damping forces from the disk cause the planets to remain close to its midplane while they migrate and the entire disk-planets system precesses rigidly very slowly around the binary. Resonant pairs are routinely formed in these simulations, mostly at the 2:1, 5:2, and 3:1 commensurabilities. In addition, the strong gravitational interaction between the resonant planets acts to reduce the impact of the binary companion after the dissipation of the disk, preserving their resonant structure, and shielding them from undergoing ZLK oscillations. Second, for closer binaries ($a_B=250$~au), the nodal precession timescale induced by the binary is in the range $10^5$--$10^6$~yr, which is similar to the disk's lifetime. In this case, the damping forces and gravitational potential from the disk are not always able to maintain the planets within its midplane, and some primordial disk-planet inclinations can be generated, especially for highly inclined binary companions. At the end of these simulations, the fraction of systems in MMRs has dropped significantly compared to simulations with $a_B=1000$~au, while the fraction of unstable systems with ejections has increased. Those ejections usually leave the remaining planets with high eccentricity. This could give an additional explanation for the existence of many S-type planets detected with high eccentricity values.

Our study shows that the existence of a binary companion can lead to significant misalignment between the planets and their natal protoplanetary disk. This suggests that misalignment between planetary orbital angular momentum and stellar spin could be generated early on in the system's lifetime, while the disk is still present. A natural continuation of this work would be to consider binary companions whose parameter distributions closely match observations, to increase the number of planets as is done in planet-planet numerical experiments, and finally to include tides and spin dynamics to properly assess the distribution of spin-orbit misalignments. 

\begin{acknowledgements}
	The authors would like to warmly thank the referee for providing insightful comments which improved this work. We also would like to warmly thank A. Crida, A. Lemaître, S. Raymond and K. Tsiganis for useful discussions. The work of A. Roisin was supported by a F.R.S.-FNRS Research Fellowship and the work of J. Teyssandier was supported by a F.R.S.-FNRS Postdoctoral Research
	Fellowship. This work was also supported by the Fonds de la Recherche Scientifique - FNRS under Grant No. F.4523.20 (DYNAMITE MIS-project). Computational resources have been provided by the Consortium des Equipements de Calcul Intensif, supported by the FNRS-FRFC, the Walloon Region, and the University of Namur (Conventions No. 2.5020.11, GEQ U.G006.15, 1610468 et RW/GEQ2016).
\end{acknowledgements}

\bibliographystyle{aa}
\bibliography{article_Roisin_Libert}

\end{document}